\documentclass[12pt,fleqn]{article}
\usepackage{amssymb,epsfig,eucal,amsmath,cite} \psfull

\numberwithin{equation}{section}

\newlength{\dinwidth}
\newlength{\dinmargin}
\def\MSbar{\overline{\mbox{\scriptsize MS}}}
\def\cO#1{{\CMcal O}\left(#1\right)}
\def\half{\mbox{\small $\frac{1}{2}$}}
\def\ho{\half\om}

\def\chieu{\chi_{{\rm eff}}^{(u)}}
\def\gae{\ga_{{\rm eff}}}
\def\nf{n_f}
\def\asb{\ab}
\def\jhep#1#2#3{{ \it JHEP } {\bf {#1}} (#3) #2}
\def\npb#1#2#3{{ \it Nucl. Phys.} {\bf {B#1}} (#3) #2}
\def\plb#1#2#3{{ \it Phys. Lett.} {\bf {#1B}} (#3) #2}

\def\pr#1#2#3{{ \it Phys. Rev.} {\bf {D#1}} (#3) #2}
\def\prep#1#2#3{{ \it Phys. Rep.} {\bf {#1}} (#3) #2}

\def\spj#1#2#3{{ \it Sov. Phys. JETP} {\bf {#1}} (#3) #2}
\def\sjnp#1#2#3{{ \it Sov. J. Nucl. Phys.} {\bf {#1}} (#3) #2}
\def\zp#1#2#3{{ \it Zeit.\ Phys.} {\bf {C#1}} (#3) #2}

\def\epj#1#2#3{{ \it Eur. Phys. J. } {\bf {C#1}} (#3) #2}
\def\ncim#1#2#3{{ \it Nuovo Cimento }{\bf {#1}} (#3) #2}
\def\hep#1{{ \tt hep-ph/#1}}

\newcommand{\ab}{\bar{\al}_\mathrm{s}}
\newcommand{\al}{\alpha}
\newcommand{\as}{\alpha_{\mathrm{s}}}                  % COSTANTE FORTE 

\newcommand{\camint}{_{{1\over2}-\ui\infty}^{{1\over2}+\ui\infty}}

\newcommand{\de}{\partial}
\newcommand{\dif}{{\rm d}}
\renewcommand{\d}{\delta}
\newcommand{\ds}{\displaystyle}
\newcommand{\De}{\Delta}

\newcommand{\esp}[1]{{\rm e}^{#1}}

\newcommand{\F}{{\CMcal F}}                  % SOLUZIONE BFKL OMOG
\newcommand{\Fr}{{\mathcal F}}              % SOLUZIONE RIDOTTA          
\newcommand{\ga}{\gamma}
\newcommand{\gb}{\bar{\gamma}}
\newcommand{\G}{{\CMcal G}}                  % FUNZIONE DI GREEN
\newcommand{\Ga}{\Gamma}
\newcommand{\Gr}{{\mathcal G}_{\om}}     % FUNZ. DI GREEN RIDOTTA

\newcommand{\K}{{\CMcal K}}                    % NUCLEO BFKL TOTALE
\newcommand{\Kr}{{\mathcal K}_{\om}}       % NUCLEO BFKL RIDOTTO
\newcommand{\kk}{{\boldsymbol k}}
\newcommand{\ku}{{\boldsymbol k}_1}
\newcommand{\kd}{{\boldsymbol k}_2}
\renewcommand{\L}{{\CMcal L}}                   % SPAZIO DI FUNZIONI

\newcommand{\La}{\Lambda}
\newcommand{\M}{{\CMcal M}}                    % AMPIEZZA
\renewcommand{\max}{{\rm Max\,}}
\renewcommand{\min}{{\rm Min\,}}
\renewcommand{\mp}{\mu_{\mathbb{P}}}     % ESTREMO DELLO SPETTRO
\newcommand{\om}{\omega}
\newcommand{\ord}{{\CMcal O}}                % O GRANDE ( DI ORDINE ...) 
\newcommand{\op}{\om_{\mathbb{P}}}           % OMEGA POMERONE

\renewcommand{\Re}{{\rm Re}}                  % PARTE REALE
\newcommand{\si}{\sigma}
\newcommand{\tb}{\bar{t}}                         % t barra
\newcommand{\Th}{\Theta}
\newcommand{\ts}{\textstyle}
\newcommand{\ui}{{\rm i}}
\newcommand{\vm}[1]{\langle#1\rangle}  % VALOR MEDIO
\newcommand{\A}{{\mathbf A}}

\begin{document}

\begin{flushright}
  hep-ph/9905566 \\
  Firenze--DFF-338-5-99\\
  Bicocca--FT--99--10\\
  May 1999
\end{flushright}
%\maketitle
\begin{center}
{\large\bf Renormalization Group Improved Small-$x$
 Equation\footnote{Work supported in part by the E.U. QCDNET
 contract FMRX-CT98-0194 and by MURST (Italy).}}
%\vspace{1cm}

M. Ciafaloni$^{\dagger}$ and D. Colferai$^{\dagger}$\\
 {\em Dipartimento di Fisica, Universit\`a di Firenze
         and INFN, Sezione di Firenze} \\
 {\em Largo E. Fermi, 2 --- 50125  Firenze}\vspace{0.5cm}\\
            G.P.  Salam$^{\dagger}$\\
 {\em Dipartimento di Fisica, Universit\`a di Milano--Bicocca
         and INFN, Sezione di Milano} \\
 {\em Via Celoria, 16 --- 20133 Milano}
\end{center}
\thispagestyle{empty}
\begin{abstract}
  We propose and analyze an improved small-$x$ equation which
  incorporates exact leading and next-to-leading BFKL kernels on one
  hand and renormalization group constraints in the relevant collinear
  limits on the other. We work out in detail the recently proposed
  $\om$-expansion of the solution, derive the Green's function
  factorization properties and discuss both the gluon anomalous
  dimension and the hard pomeron. The resummed results are stable,
  nearly renormalization-scheme independent, and join smoothly with
  the fixed order perturbative regime. Two critical hard pomeron
  exponents $\om_c(Q^2)$ and $\om_s(Q^2)$ are provided, which --- for
  reasonable strong-coupling extrapolations --- are argued to provide
  bounds on the pomeron intercept $\op$.
\end{abstract}
\begin{center}
\end{center}
\vspace*{0.5 cm}
{\small $~^{\dagger}$ e-mail: ciafaloni@fi.infn.it,
  colferai@fi.infn.it, gavin.salam@mi.infn.it}
\newpage

%========================= 1 ========================

\section{Introduction}

Recent results on the next-to-leading $\log s$
corrections~\cite{fl98,cc98} to the BFKL equation~\cite{bfkl} and to
the hard pomeron show subleading effects which are so large as to
question the very meaning of the high-energy expansion and thus raise
the compelling question of how to improve it.

Two facts suggest that an essential ingredient of any improvement of
the BFKL approach should be the correct treatment of the collinear
behavior, as predicted by the renormalization group (RG): on the one
hand the success of normal QCD evolution~\cite{dglap} in explaining
the $Q^2$ dependence of the small-$x$ behavior of structure functions
at HERA, and on the other hand the observation that the large
next-to-leading corrections to the BFKL equation come mostly from
collinearly-enhanced physical contributions.

A first attempt at introducing collinear improvements was performed
long ago, by the treatment of coherence effects~\cite{c88} in the
collinear region. This leads to the CCFM equation~\cite{c88,cfm90},
which differs from the BFKL equation by subleading effects to all
orders, even if a full inclusion of the DGLAP splitting functions in a
consistent CCFM framework has not yet been achieved. Other modified
BFKL approaches incorporating some DGLAP evolution are being tried
too~\cite{kms97}.

Very recently, it has been suggested~\cite{cd99} that such all-order
collinear effects can be incorporated as subleading kernels of a
generalized equation, whose solution can be found by the method of the
so called $\om$-expansion, allowing in particular a resummation of the
energy-scale-dependent terms of the kernel~\cite{s98}.

The purpose of the present paper is to insert such suggestions in a
general scheme, which leads to the renormalization group improved
small-$x$ equation, and to study its solutions. One of the outcomes
will be to stabilize, in a nearly scheme-independent way, the estimate
of the anomalous dimensions and of the $Q^2$-dependent hard pomeron.

A first point to clear up is which pomeron we are going to
estimate. Previous work on RG factorization~\cite{ck89,cc97} in the
BFKL equation with running
coupling~\cite{glr,l86,hr92,nz94,l95,b95,hkk97,km98} has shown that
the {\em pomeron} $\op$ defined as the $Q^2$-independent leading
singularity in the $\om$-plane, is beyond the reach of a strictly
perturbative approach. On the other hand, there appears to be a
boundary of validity of the RG, the $Q^2$-dependent {\em hard pomeron}
$\op(Q^2)$, which is argued to be {\em independent} of the
small-$\kk^2$ strong coupling region and is thus hopefully calculable
in perturbative QCD.

Since $\op(Q^2)$ signals a change of asymptotic regime, it is
associated with an $\om$-singularity of the anomalous dimensions, not
necessarily of the full gluon distribution. Thus $\op(Q^2)$ may be
related to a power-like behavior in an intermediate small-$x$,
moderate $Q^2$ region, and not to the very small-$x$ asymptotic
behavior of the structure functions. It also follows that $\op(Q^2)$
is a rather difficult quantity to determine, because it is related to
the position of an $\om$-singularity, and is thus dependent on the
full anomalous dimension perturbative series. Possible definitions,
leading to a precise estimate, are thoroughly discussed in
Sec.~\ref{s:sxefgk} and in Sec.~\ref{s:adhp}, where our results are
provided. 

A second point to realize is that, in order to incorporate the
collinear behavior correctly, a whole string of subleading kernels,
represented by a series in the running coupling $\as(Q^2)$, is to be
taken into account.  In fact, the leading $\log s$ calculations count
one high-energy gluon exchange per power of $\as$, with any transverse
momentum ratios. In the collinear limit, provided by the strong
ordering in the transverse momenta, only the singular part $\sim1/z$
of the DGLAP splitting function is obtained.  The remaining part
contributes to higher and higher order subleading kernels which carry
fewer powers of $\log s$, but leading powers of $\log Q^2$.
From a quantitative point of view, such collinear contributions are
very important, and in fact account for most of the exact NL kernel
(Sec.~3).

Let us stress that we are not aiming here at a full control of
subleading $\log s$ contributions, but only of those that carry a
leading collinear contribution from NNL level on. Therefore, we remain
in a context in which only $t$-channel iteration is important, without
mixing with the $s$-channel one (see, e.g.,~\cite{glr,bkp80}).

In this framework, we can define the $\kk$-dependent gluon
distribution by the NL $\kk$-factorization formula introduced by one
of us~\cite{c98} in large $\kk$ dijet production in parton-parton
scattering.  In a general hard process involving probes $A$ and $B$ we
can write~\cite{cc98}
\begin{align}
 {\dif\si_{AB}\over\dif\kk\;\dif\kk_0}=\int{\dif\om\over2\pi\ui}
 \left({s\over kk_0}\right)^{\om}&h_A(\kk)\langle\kk|\big(1+\as H\big)
 \;\G_{\om}\;\big(1+\as H^{\dagger}\big)|\kk_0\rangle
 h_B(\kk_0)\,,            \label{fatt}\\
 &k=|\kk|\,,\qquad k_0=|\kk_0|  \nonumber
\end{align}
where the impact factors $h_A$, $h_B$ may carry additional dependence on
the hard scales of the probes and the gluon Green's function is provided by
\begin{equation}
 \G_{\om}(\kk,\kk_0)=\langle\kk|[\om-\K_{\om}]^{-1}|\kk_0\rangle
 \,, \label{defg}      
\end{equation}
apart from the multiplicative kernels $H$, $H^{\dagger}$ which may be
needed at subleading level~\cite{cc98,cd98}.

We notice immediately that the scale of the energy $s$ in
Eq.~(\ref{fatt}) has been chosen to be $kk_0$, i.e., factorized and
symmetrical in the ``upper'' ($u$) scale $k$ and ``lower''
($l$) scale $k_0$. This means that $\G_{\om}$ and the kernel
$\K_{\om}$ in Eq.~(\ref{defg}) are both symmetrical operators. On the
other hand, when $k\gg k_0$ ($k_0\gg k$), the variable $kk_0/s$ {\em
is not} the correct scaling variable, which is rather $k^2/s$
($k_0^2/s$) --- i.e.  the usual Bjorken variable.

In order to switch to, say, the upper energy-scale $k^2$, it is
apparent from Eq.~(\ref{fatt}) that one has to perform a similarity
transformation $\G_{\om}\to(k/k_0)^{\om}\G_{\om}$, which in turn
implies the relationship
\begin{equation}
 \K_{\om}^{(u)}(\kk,\kk')\left(k'\over k\right)^{\om}=
 \K_{\om}(\kk,\kk')=
 \K_{\om}^{(l)}(\kk,\kk')\left(k\over k'\right)^{\om}  \label{cambioscala}
\end{equation}
between the symmetrical kernel $\K_{\om}$ and the kernel $\K^{(u)}$
($\K^{(l)}$). Although technical, this remark is important in order to
classify the collinear logarithms, because if a wrong energy-scale is
chosen, single logs (of $k/k'$) may turn into double logs (cf.\ Sec.~3).

The main purpose of our study is to construct the RG improved kernel
$\K_{\om}$, and to provide the solution for $\G_{\om}$ in
Eq.~(\ref{defg}) in the RG regime $k^2\gg k_0^2\gg\La^2$. The starting
point is the observation~\cite{cd99} that the kernel
$\K_{\om}(\kk,\kk';\mu^2;\as(\mu^2))$, for non-vanishing values of
$\kk,\kk'$, is RG invariant, and can thus be expanded as a power
series in $\as(k^2)$ with scale invariant coefficients
\begin{equation}
 \K_{\om}(\kk,\kk')=\sum_{n=0}^{\infty}[\ab(k^2)]^{n+1}\;
 K_n^{\om}(\kk,\kk')\,,\qquad\ab={N_c\as\over\pi}=
 {1\over b\log(k^2/\La^2)}\,.           \label{serie}
\end{equation}
Since we want to take into account the leading collinear singularities
to all-orders, the series (\ref{serie}) is necessarily infinite, as
noticed before.

Solving for the Green's function with the general kernel (\ref{serie})
is a novel problem in the BFKL approach, which is addressed and solved
in Sec.~\ref{s:sxefgk}.  There we derive the main properties of the
solutions, namely ($i$) the factorization property of $\G_{\om}$ in
the RG regime, ($ii$) the $\om$- expansion of the relevant
eigenfunctions and ($iii$) the definitions of the pomeron singularity
$\op$ and of the hard pomeron singularity $\op(\as(k^2))$.  In a first
reading of this rather mathematical section one could perhaps retain
the basic results, and come back to their derivations after Secs.~3-5.

In Sec.~3 we explicitly construct the improved kernel $\K_{\om}$ with
the requirements of ($i$) reducing to the exact L+NL terms in the
relevant limit and ($ii$) reproducing the known collinear
singularities at higher orders.

The corresponding solution for $\G_{\om}$ in the RG regime and the
explicit form of the solution $\F_{\om}$ of the homogeneous equation
are studied in Sec.~4. The main result is that the NL truncation of
the improved $\om$-expansion takes into account correctly all
collinear singularities, at least for the purely gluonic case.  The
inclusion of the (small) $q\bar{q}$ contributions is discussed also.

Finally, in Sec.~\ref{s:adhp} we present our results for the resummed
anomalous dimensions and for the hard pomeron, and exhibit their
stability under scheme change and NNL corrections.

In the last Sec.~\ref{sec:concl} we discuss the present situation and
future prospects, which include a solvable model~\cite{cds99b}, based
on the collinear analysis of the present paper. A few mathematical
details are covered in the Appendix.

%========================= 2 ========================

\section{Small-$x$ equation for a general kernel}
\label{s:sxefgk}

We consider here a general form of the small-$x$ equation, whose
$\as(t)$-dependence is supposed to be consistent with leading-twist
anomalous dimensions and must contain, therefore, an infinite series
of subleading terms (cf.\ Introduction). Our final goal is to
investigate the solution for the gluon Green's function (\ref{defg}),
i.e., the resolvent of the improved kernel
\begin{equation}
 \om\G_{\om}(\kk,\kk_0)=\d^2(\kk-\kk_0)+\int{\dif^2\kk'\over\pi}\;
 \K_{\om}(\kk,\kk')\G_{\om}(\kk',\kk_0)\,,     \label{g}
\end{equation}
in order to derive its large-$t$ behavior in the RG regime.

\subsection{Form of the kernel}

The improved kernel $\K_{\om}(\kk,\kk')$ occurring in Eq.~(\ref{g}) is
assumed to have the asymptotic $\as(t)$-expansion
\begin{equation}
 \K_{\om}(\kk,\kk')=\sum_{n=0}^{\infty}[\ab(t)]^{n+1}\;
 K_n^{\om}(\kk,\kk') \,,\qquad t\equiv\log{k^2\over\La^2}\label{seriet}
\end{equation}
where the coefficient kernels $K_n^{\om}$ are scale-invariant and may
be $\om$-dependent. They are partly known in closed form from
leading~\cite{bfkl} and next-to-leading~\cite{fl98,cc98} calculations,
and have known~\cite{cd99} collinear properties to all orders.

The leading coefficient kernel $K_0^{\om}$ must reduce, for $\om\to0$, 
to the historical one~\cite{bfkl} having eigenvalue function
\begin{equation}
 \chi_0(\ga)=2\psi(1)-\psi(\ga)-\psi(1-\ga)\,,\qquad
 \psi\equiv{\Ga'\over\Ga}\,,\label{chi0}
\end{equation}
on test functions $(k^2)^{\ga-1}$. The NL coefficient kernel
$K_1^{\om}$ is related also, for $\om\to0$, to the one recently
found~\cite{fl98,cc98} on the basis of NL QCD vertices, except
for the subtraction of a term already included in the $\om$-dependence
of $K_0^{\om}$ (cf.\ Sec.~3).

In general, the expansion (\ref{seriet}) was justified in
Ref.~\cite{cd99} as follows. First $\K_{\om}(\kk,\kk')$, at
energy-scale $s_0=kk_0$ (Eq.~(\ref{fatt})) and non vanishing
virtualities, is a collinear finite distribution, symmetrical in its
arguments. By RG equations, for $k$ and $k'$ much larger than $\La$,
$\K_{\om}$ must have the form
\begin{equation}
 \K_{\om}(\kk,\kk';\mu^2;\as(\mu^2))=
 {\ab(t)\over k^2}\hat{\K}_{\om}(\ab(t);t,t')=
 {\ab(t')\over k^{'\,2}}\hat{\K}_{\om}(\ab(t');t',t)  \label{tetto}
\end{equation} 
which, by expanding in $\ab(t)$, yields Eq.~(\ref{seriet}).

In the limit of vanishing virtualities ($k\to0$ or $ k'\to0$)
$\K_{\om}$ acquires collinear singularities, which are
dictated by the nonsingular part of the gluon anomalous
dimension in the $Q_0$-scheme which, by neglecting the $q\bar{q}$
part, is
\begin{align}
 \tilde{\ga}(\om)&=\ga_{gg}(\om)-{\ab\over\om}=\ab A_1(\om)+
 \ab^2A_2(\om)+\cdots\,,       \label{dimanom}\\
 A_1(\om)&=-{11\over12}+\ord(\om)
 \,,\qquad A_2(\om)=0+\ord(\om)\,.\nonumber
\end{align}
As a consequence, the eigenvalue functions $\chi_n^{\om}(\ga)$
acquire the $\ga$-singularities
\begin{align}
 \chi^{\om}_n(\ga)&\simeq{1\cdot A_1(A_1+b)\cdots(A_1+(n-1)b)\over
 (\ga+\ts{1\over2}\om)^{n+1}}\,,\qquad(\ga\ll1)\nonumber\\
 &\simeq{1\cdot(A_1-b)(A_1-2b)\cdots(A_1-nb)\over
 (1-\ga+\ts{1\over2}\om)^{n+1}}\,,\qquad(1-\ga\ll1)\,,\label{chin}
\end{align}
where $b$ is the one-loop beta function coefficient (cf.\ Sec.~3).

The $\ga$, $\om$ dependences are tied up together in
Eq.~(\ref{chin}) because of the similarity relations
(\ref{cambioscala}), which define the kernels $\K^{(u)}$ ($\K^{(l)}$) at
energy-scale $k^2$ ($k_0^2$) having simple collinear behavior for
$k\gg k'$ ($k'\gg k$). As a consequence, the $\ga$-singularities
occur at shifted values of $\ga$ (by $\pm\om/2$) and the symmetry of
$\K_{\om}$ implies, by Eq.~(\ref{seriet}), a slightly asymmetrical
$b$-dependence in Eq.~(\ref{chin}).

\subsection{Factorization of non-perturbative effects}

In order to actually solve Eq.~(\ref{g}) for $\G_{\om}(\kk,\kk_0)$, one
should extend the representation (\ref{seriet}) in the region around
the Landau pole $k\simeq\La$ ($t=0$), where it becomes unreliable.
Whether such an extension can be somehow hinted at on perturbative
grounds --- as in the time-like evolution case~\cite{dmw96} --- is an open
problem that we do not address here. However, for perturbation theory
to be applicable, the non-perturbative effects of such region should
be factorized out, as is predicted by the RG, and has been argued for
at leading and first subleading level~\cite{glr,cc97}.

In the following, we consider the dependence of $\G_{\om}$ on various
kinds of regularization of $\K_{\om}$ in Eq.~(\ref{seriet}) around the
Landau pole, and we argue that indeed the RG factorization property
holds for sufficiently large $t$, in the form
\begin{equation}
 \G_{\om}(\kk,\kk_0)=\F_{\om}(\kk)\tilde{\F}_{\om}(\kk_0)+\text{higher
 twist terms}\,,\qquad(t-t_0\gg1)\label{fattg}\,.
\end{equation}
Here $\F_{\om}$ ($\tilde{\F}_{\om}$) is the solution of the
homogeneous small-$x$ equation
\begin{equation}
 \K_{\om}\F_{\om}=\om\F_{\om}\,,\label{eqbfkl}
\end{equation}
which is ``regular'' for $t\to+\infty$ ($t\to-\infty$) in the sense
that it is asymptotically $\L^2$ in the corresponding region (see
Sec.~\ref{ss:fots}  for a more precise discussion).

Let us first try to understand how Eq.~(\ref{fattg}) can possibly
work. By inserting it in the defining equations
\begin{subequations}
\begin{align}
 \om\Gr(t,t_0)&-\int\dif t'\;\Kr(t,t')\,\Gr(t',t_0)=\d(t-t_0)\,,\nonumber\\
 &\Kr(t,t')\equiv kk'\K_{\om}(\kk,\kk')\,,\qquad
 \Gr(t,t_0)\equiv kk_0\G_{\om}(\kk,\kk')\,, \label{ridotte}
\end{align}
and  by using the symmetry of $\Gr$, we obtain, for $t-t_0\gg1$,
\begin{align}
 \om\Fr_{\om}(t)&-\int_{-\infty}^{+\infty}\dif t'\;\Kr(t,t')\,\Fr_{\om}(t')
 \simeq    \label{efferidotte}\\
 &\simeq\int_{-\infty}^{t_0}\dif t'\,\Kr(t,t')\left[\tilde{\Fr}_{\om}(t')\,
 {\Fr_{\om}(t_0)\over\tilde{\Fr}_{\om}(t_0)}-\Fr_{\om}(t')\right]-
 \int_{-\infty}^{+\infty}\dif t'\,\Kr(t,t'){\De_{\om}(t',t_0)\over
 \tilde{\Fr}_{\om}(t_0)}\,,       \nonumber\\
 &\Fr_{\om}(t)\equiv k\F_{\om}(\kk)\,,\qquad
 \tilde{\Fr}_{\om}(t)\equiv k\tilde{\F}_{\om}(\kk)\,, \nonumber
\end{align}
\end{subequations}
where $\De_{\om}(t,t_0)$ denotes the higher twist part of $\G_{\om}$
in Eq.~(\ref{fattg}).  Now, let us go to the large-$t$ limit: the
l.h.s. is the homogeneous small-$x$ equation for $\Fr_{\om}$, and the
r.h.s. will be negligible, i.e., higher twist, by the following
mechanism.  First, note that $t'\lesssim\cO{t_0}$ in the r.h.s.,
because $\De_{\om}$, by definition, decreases rapidly for
$|t'-t_0|\gg1$. Furthermore, for $|t-t'|\gg1$, $\K_{\om}$ satisfies
the collinear factorization of Sec.~3 (with higher twist corrections),
so that the $t_0$-dependence in the r.h.s. is factored out and can be
made to vanish by a proper choice of $\De_{\om}$.

We thus conclude that, provided the regularization of the running
coupling allows such properties of the kernel, the factorization in
Eq.~(\ref{fattg}) of the large-$t$ dependence actually holds. The
decomposition of the kernel in a factorizable and in a local part is
certainly satisfied in the case of models leading to differential
equations (cf.\ Ref.~\cite{cc97} and the collinear model of
Ref.~\cite{cds99b} as soluble examples), but is presumably satisfied
also in the case of kernels in an $\L^2$ space having reasonable
spectral properties, as we shall argue next.

\subsection{Form of the solution \label{ss:fots}}

We thus assume that, by a suitable regularization of $\as(t)$ around
the Landau pole, $\K_{\om}$ can be defined as a hermitian operator
bounded from above in an $\L^2$ Hilbert space, with a continuum (or
possibly discrete) spectrum $-\infty<\mu<\mp(\om)$. Typical
regularizations of this kind may (a) cut-off $\as(t)$ below some value 
$t=\tb>0$, or (b) freeze it in the form
$\ab(t)=(bt)^{-1}\Th(t-\tb)+(b\tb)^{-1}\Th(\tb-t)$, possibly with some 
smoothing out around the cusp. The spectrum of $\K_{\om}$ is expected
to be discrete in case (a) and continuum in case (b)~\cite{cc97}. In the
latter case, the expansion in Eq.~(\ref{seriet}), extended to the
region $t<\tb$, defines a scale-invariant kernel with frozen coupling,
where however the coefficient kernels $K_n^{\om}$ should be evaluated, 
for consistency, in the $b=0$ limit. This limit introduces some
ambiguity in the definition of $K_n^{\om}$ below $\tb$, which in our
point of view is part of the regularization procedure.

In such a framework, a formal solution for the Green's function
$\G_{\om}$ is given by the spectral representation
\begin{equation}
 \G_{\om}(\kk,\kk_0)=\int_{-\infty}^{\mp(\om)}{\dif\mu\over\pi}\;
 {\F_{\om}^{\mu}(\kk)\F_{\om}^{\mu\,*}(\kk_0)\over\om-\mu}\label{rapspet}
\end{equation}
in terms of the eigenfunctions
\begin{equation}
 \K_{\om}\F_{\om}^{\mu}=\mu\F_{\om}^{\mu}\,,\label{autof}
\end{equation}
which satisfy an $\L^2$ orthonormality condition
\begin{equation}
 (\F_{\om}^{\mu},\F_{\om}^{\mu'})\equiv\int{\dif^2\kk\over\pi}\;
 \F_{\om}^{\mu\,*}(\kk)\F_{\om}^{\mu'}(\kk)=\int\dif t\;
 \Fr_{\om}^{\mu\,*}(t)\Fr_{\om}^{\mu'}(t)=\d(\mu-\mu')\quad\label{orto}
\end{equation}
and can be chosen to be real, because so is $\K_{\om}(\kk,\kk')$.

We shall normally consider the situation for which $\Re(\om)>\mp(\om)$,
so that $\om$ is not a point of the spectrum (\ref{autof}), and
$\F_{\om}$ ($\tilde{\F}_{\om}$) in Eqs.~(\ref{fattg}) and
(\ref{eqbfkl}) are not eigenfunctions, being well behaved for
$t\to+\infty$ ($t\to-\infty$) only.

We shall also refer, in most of this section, to the example of the
frozen-$\as$ regularization, which allows a simple classification of
the eigenfunctions $\F_{\om}^{\mu}(\kk)$ of Eq.~(\ref{autof}), according 
to their behavior for $t\to-\infty$. In fact, since the test functions
\begin{equation}
 (k^2)^{-\ga(\mu)}=(k^2)^{-{1\over2}}\esp{\ui\nu(\mu)t}\,,
 \quad(1-\ga=1/2+\ui\nu)\,,       \label{afz}
\end{equation}
are reproduced for large negative $t$ by the kernel (\ref{seriet})
with eigenvalues
\begin{equation}
 \mu=\sum_{n=0}^{\infty}[\ab(\tb)]^{n+1}\chi_n^{\om}(1/2+\ui
 \nu(\mu),b=0)           \label{avl}
\end{equation}
the eigenfunctions $\F_{\om}^{\mu}(\kk)$ must have the behavior
\begin{align}
 \F^{\mu}(\kk)&\underset{\qquad\;}{=}\;{1\over2\ui}\left(F^{\nu(\mu)}(\kk)-
 F^{-\nu(\mu)}(\kk)\right)   \label{onde}\\
 &\underset{t\to-\infty}{\simeq}\;{1\over2\ui k}\left(\tau(\nu)\esp{\ui
 \nu(\mu)t}-\tau^*(\nu)\esp{-\ui\nu(\mu)t}\right)\nonumber
\end{align}
for suitable functions $F^{\nu}(\kk)$ having a plane-wave asymptotic
behavior for large and negative $t$ (the $\om$ index has been
dropped).  The two ``frequencies'' $\nu(\mu)$ and $-\nu(\mu)$
correspond to the two solutions of Eq.~(\ref{avl}) for real $\mu$,
which are real also, because of the $\ga\leftrightarrow1-\ga$ symmetry
of $\chi_n^{\om}(\ga)$ in the $b=0$ limit, as better seen from their
explicit form, similar to the basic one in Eq.~(\ref{chi0})
(Sec.~3). Note also that the spectrum endpoint is provided by the
maximum of the (real) expression (\ref{avl}) when $\nu$ varies.

The precise superposition of left- and right-moving waves occurring in 
Eq.~(\ref{onde}) is determined by the condition that $\F^{\mu}(\kk)$ be
regular for $t\to+\infty$, i.e., be vanishing at least as rapidly as
$1/k$, so as to allow an $\L^2$ (continuum) normalization. While the
negative-$t$ behavior (\ref{onde}) is oscillating for $\mu$ in the
spectrum (\ref{avl}), it becomes a superposition of decreasing and
increasing exponentials when $\mu$ is continued off the real axis with 
$\Re(\ui\nu)>0$. This structure, similar to that of potential
scattering~\cite{cc97}, suggests that the Green's function can be
asymptotically evaluated by the ``on shell'' expression
\begin{equation}
 \G_{\om}(\kk,\kk_0)\simeq\F_{\om}^{\om}(\kk)\,F_{\om}^{\nu(\om)}(\kk_0)\,,
 \qquad t-t_0\gg1\,,     \label{fattor}
\end{equation}
thus identifying $\tilde{\F}_{\om}(\kk_0)=F_{\om}^{\nu(\om)}(\kk_0)$ in
Eq.~(\ref{fattg}) as the solution of the homogeneous BFKL equation
which is regular for $t_0\to-\infty$.

The argument goes as follows. By using Eq.~(\ref{onde}) we rewrite the 
spectral representation (\ref{rapspet}) as a contour integral
\begin{equation}
 \G_{\om}(\kk,\kk_0)=\int_{C(\om)}{\dif\mu\over2\pi\ui}\;
 {\F_{\om}^{\mu}(\kk)F_{\om}^{\nu(\mu)}(\kk_0)\over\om-\mu}\,,
 \label{intcont}
\end{equation}
where $F^{\nu(\mu)}$ and $F^{-\nu(\mu)}$ are assumed to be boundary
values of an imaginary analytic function of $\mu$, whose branch cut
lies along the spectrum and is encircled by the contour $C(\om)$. We
then evaluate the behavior of (\ref{intcont}) for $t_0\to-\infty$, by
distorting the $\mu$-contour (because $F^{\nu(\mu)}$ is well behaved,
for $\Re(\ui\nu)>0$) and by picking up the residue at the $\mu=\om$
pole, i.e., the r.h.s. of Eq.~(\ref{fattor}). This procedure can be
carried through for $t_0>0$ also, where $F^{\nu(\om)}$ becomes
irregular, provided $t-t_0$ is large enough for the decrease of
$\F^{\om}$ to compensate the increase of $F^{\nu(\om)}$.

The plausibility argument above is further supported by the explicit
model of Ref.~\cite{cds99b} for arbitrary values of $t$ and $t_0$, and
hints at the general validity of Eq.~(\ref{fattor}). Therefore, for
$k\gg k_0$, the Green's function is asymptotically proportional to the
``on shell'' regular solution of the homogeneous BFKL equation
$\F_{\om}^{\om}(\kk)$, which becomes the basic quantity to be found.

Furthermore, the above procedure allows us to define also the pomeron
singularity $\om=\op$. In fact, the integral representation
(\ref{intcont}) is singular when its contour is pinched between the
branch-point $\mu=\mp(\om)$ and the pole $\mu=\om$, i.e., for
$\nu(\mp=\op)=0$, or
\begin{equation}
 \op=\mp(\op)=\sum_{n=0}^{\infty}[\ab(\tb)]^{n+1}\chi_n^{\op}(1/2
 ,b=0)           \label{omegap}
\end{equation}
which is an implicit equation for $\op$ in the present regularization
procedure of $\as$-freezing at small $\kk$. For a general
regularization, the definition $\op=\mp(\op)$ is still valid, but the
explicit expression (\ref{omegap}) is not.

It follows that $\op$ is a singularity of the right-moving wave
$F_{\om}^{\nu(\om)}$ rather than the regular solution, and that
it affects the asymptotic behavior (\ref{fattor}) in the
$t_0$-dependent coefficient only. Therefore, the regularization
dependence of $\op$ and of the spectrum is factorized away
asymptotically. This picture is confirmed by the explicit examples of
Refs.~\cite{cc97,cds99b}.

\subsection{Small-$\om$ expansion}\label{s-oe}

We follow the philosophy of Ref.~\cite{cd99}, according to which
$\om\ll1$ is the relevant expansion parameter of the solution, rather
than $\as(t)$. Furthermore, we first consider the ``off-shell'' case
$\mu\neq\om$, or more precisely $\mu\leq\mp(\om)<\Re(\om)\ll1$,
and we take the generalized ansatz
\begin{equation}
 f_{\om}^{\mu}(t)\equiv k^2\F_{\om}^{\mu}(\kk)=\int\camint{\dif\ga\over
 2\pi\ui}\exp\left\{\ga t-{1\over b\mu}X_\om(\ga,\mu)\right\}\;\;
 ,\;\; b={\pi\over N_c}\left({11N_c-2 \nf\over12\pi}\right)
 \label{rappres}
\end{equation}
where $X_\om(\ga,\mu)$ is to be found by solving Eq.~(\ref{autof}).

Once again, we are interested in the RG regime $bt\gtrsim1/\mu\gg1$,
in which the regular solution in Eq.~(\ref{rappres}) turns out to be
dominated by the stable saddle point $\ga=\gb_{\om}(\mu,t)$ defined by
\begin{align}
 \de_{\ga}\left\{\ga t-{1\over b\mu}X_\om(\ga,\mu)
 \right\}_{\ga=\gb}=0&\quad\iff\quad b\mu t=\chi_\om(\gb,\mu)\equiv
 X'_\om(\gb,\mu)\,,\nonumber\\
 &-\chi'_\om(\gb,\mu)>0\,,\qquad
 (\#)'\equiv{\de\over\de\ga}(\#)\,.       \label{puntos}
\end{align}
It has already been shown~\cite{cd99} that, around the saddle point, the
effective eigenvalue function $\chi_{\om}(\ga,\mu)$ is independent of
the regularization procedure and its $\mu$-expansion has been
evaluated for $\mu=\om$, by a treatment of the saddle point
fluctuations (App.~\ref{a:puntosella}).

Here we prefer to find the $\mu$-expansion, in the same regime, by
using the replacement $t\to\de_{\ga}$ in $\ga$-space~\cite{ck89,glr}, in
order to give an all-order evaluation. We thus write Eq.~(\ref{autof})
for $t>\tb$ in the form
\begin{equation}
 b\mu t\,f_{\om}^{\mu}=\left(\hat{\K}_0+{1\over b\mu t}\,\mu\,
 \hat{\K}_1+{1\over(b\mu t)^2}\,\mu^2\,\hat{\K}_2+\cdots\right)
 f_{\om}^{\mu}           \label{svileq}
\end{equation}
and by repeated partial integrations we prove the $\ga$-space identity
\begin{equation}
 b\mu\hat{t}\big(g(\ga){f}_{\om}^{\mu}(\ga)\big)=\left[\left(
 \chi_{\om}(\ga,\mu)-b\mu\de_{\ga}\right)g(\ga)\right]
 {f}_{\om}^{\mu}(\ga)\,.        \label{gid}
\end{equation}

Strictly speaking, the validity of Eq.~(\ref{gid}) is limited by the
fact that the $b\mu\hat{t}$ operator has to be regularized around
$t=0$ (e.g., by freezing it for $t<\tb$). However, the large-$t$
behavior of Eq.~(\ref{rappres}) can be safely evaluated by (\ref{gid}) 
provided
\begin{equation}
 b\mu t\simeq\chi_{\om}(\gb,\mu)\gg b\mu\tb\,,  \label{adr}
\end{equation}
by the saddle point condition (\ref{puntos}).

By replacing (\ref{gid}) into (\ref{svileq}) we obtain the equation
\begin{equation}
 \chi_{\om}(\ga,\mu)=\chi_0^{\om}(\ga)+\left(\chi^{\om}-b\mu\de_{\ga}
 \right)^{-1}\mu\chi_1^{\om}(\ga)+\left(\chi^{\om}-b\mu\de_{\ga}
 \right)^{-2}\mu^2\chi_2^{\om}(\ga)+\cdots    \label{equazdiff}
\end{equation}
which, at a given subleading order in $\chi^{\om}_n$ provides a
nonlinear differential equation for $\chi_{\om}(\ga,\mu)$, and thus a
formal solution of Eq.~(\ref{svileq}).

However, since we are looking at the large-$t$ and small-$\mu$ limits, 
we prefer to expand Eq.~(\ref{equazdiff}) in the denominators as well, 
thus obtaining the following asymptotic expansion:
\begin{equation}
 \chi_{\om}(\ga,\mu)=\chi^{\om}_0(\ga)+\mu\,\eta_1^{\om}(\ga)+\mu^2
 \,\eta_2^{\om}(\ga)+\cdots\,,       \label{omegaes}
\end{equation}
where (App.~\ref{a:dergamma})
\begin{align}
 \eta_1^{\om}&={\chi_1^{\om}\over\chi_0^{\om}}\,,\nonumber\\
 \eta_2^{\om}&={1\over\chi_0^{\om}}\left[{\chi_2^{\om}\over\chi_0^{\om}
 }+b\left({\chi_1^{\om}\over\chi_0^{\om}}\right)^{\prime}-\left({
 \chi_1^{\om}\over\chi_0^{\om}}\right)^2\right]\,,   \label{ete}\\
 \eta_3^{\om}&={1\over\chi_0^{\om}}\left[{\chi_3^{\om}\over(\chi_0^{\om}
 )^2}+{b\over\chi_0^{\om}}\left({\chi_2^{\om}\over\chi_0^{\om}}\right)
 ^{\prime}-{\chi_1^{\om}\chi_2^{\om}\over(\chi_0^{\om})^3}+b\,\eta_2
 ^{\om}{}'-2\,\eta_1^{\om}\eta_2^{\om}\right]\,,\nonumber
\end{align}
and so on. This expansion is supposed to yield safely the large-$t$
behavior of Eq.~(\ref{rappres}), whenever Eq.~(\ref{adr}) is
satisfied. The cumbersome saddle point fluctuation method of
App.~\ref{a:puntosella} checks with the result in Eq.~(\ref{ete}).

The $\mu$-expansion of the regular solution in
Eqs.~(\ref{omegaes},\ref{ete}) is the basic result of this section and 
will be applied in the following ones to actual NL calculations.

\subsection{Anomalous dimension and hard pomeron}

Due to the validity of RG factorization in the large-$t$ limit of
Eq.~(\ref{fattg}), we can state that the gluon density $g_{\om}^A(t)$
in the probe $A$ has a universal $t$-dependence
\begin{equation}
 \dot{g}_{\om}^A(t)={\de\over\de t}g_{\om}^A(t)\sim k^2\G_\om
 (\kk,\kk_0)\sim k^2\F_{\om}(\kk)\,C_{\om}^A\,,      \label{densita}
\end{equation}
where the $A$-dependent coefficient is in general non-perturbative. By 
Eq.~(\ref{rappres}), this yields the proportionality relation
\begin{align}
 g_{\om}^A(t)&=g_{\om}(t)\,C_{\om}^A\,,  \nonumber\\
 g_{\om}(t)&=\int{\dif\ga\over2\pi\ui}\;{1\over\ga}\,\exp
 \left\{\ga t-{1\over b\om}X^{(u)}_{\om}(\ga,\om)\right\}\,,\label{gint}
\end{align}
where we have specified the $X_{(u)}^{\om}$ function at the ``upper''
energy-scale $s_0=k^2$ ($\ga\to\ga-{1\over2}\om$), which is relevant
in the large $k^2$ limit. 

The asymptotic behavior of Eq.~(\ref{gint}) in the RG regime can be
found from the saddle point (\ref{puntos}), which yields the result
\begin{equation}
 g_{\om}(t)\simeq\left({1\over\gb(\om,t)\sqrt{-\chi^{(u)}{}'(\gb,\om)}}+
 \cdots\right)\exp\int^t\gb(\om,\tau)\;\dif\tau\,,  \label{gdiman}
\end{equation}
where $\gb(\om,t)\equiv\gb_{\om}(\om,t)$ satisfies the identity
\begin{equation}
 \gb t-X^{(u)}_{\om}(\gb,\om)=\int^t\gb(\om,\tau)\;\dif\tau+
 {\rm const}\label{espgb}
\end{equation}
and the coefficient in front, coming from the saddle point
fluctuations, has been evaluated at NL level only.

If we work at NL level, the saddle point approximation (\ref{gdiman})
is enough, and provides the effective anomalous dimension~\cite{cc97b}
\begin{equation}
 \ga_{{\rm eff}}(\om,t)=\gb(\om,t)-{b\om\over\chi^{(u)}{}'(\gb,\om)}
 \left({1\over\gb}+{1\over2}{\chi^{(u)}{}''(\gb,\om)\over\chi^{(u)}{}'
 (\gb,\om)}\right)+\cdots\label{gaeff}
\end{equation}
whose subleading expansion has however a wildly oscillating
behavior~\cite{bv98,r98}. The {\em hard pomeron} singularity comes in
this case from the failure of the saddle point expansion at the point
$\om=\om_{s}(t)$), such that
\begin{equation}
 \chi'(\gb(\om_{s},t),\om_s)=0\qquad
 \text{(saddle point estimate)}\,,\label{omsp}
\end{equation}
thus implying infinite fluctuations in Eq.~(\ref{gaeff}).

On the other hand, in our RG improved approach, we do not rely on a
subleading hierarchy. Therefore, the estimate (\ref{omsp}) may be not 
realistic. For instance, it has been suggested~\cite{t99} that
Eq.~(\ref{gaeff}) yields higher-order singularities of oscillating
type which may perhaps resum to a scale change. Here we just notice
that Eq.~(\ref{gint}), with the solution (\ref{omegaes}) can simply be 
evaluated beyond the saddle point approximation for $\om\leq\om_{s}$, 
and yields a generalized definition of the effective anomalous dimension
\begin{equation}
 \ga_{{\rm eff}}(\om,t)={\dot{g}_{\om}^A(t)\over g_{\om}^A(t)}=
 {k^2\F_{\om}(\kk)\over g_{\om}(t)}\,.  \label{derlog}
\end{equation}

The analysis of the $\om$-singularities of Eq.~(\ref{derlog}) has been
done in some toy models in which the BFKL equation reduces to a
differential one~\cite{cc97,cds99b}. In such cases the singularity
comes just from a zero of the denominator at $\om=\om_{c}(t)$, so that
\begin{equation}
 g_{\om_c(t)}(t)=0\qquad\text{(gluon-density-zero estimate)}\,.
 \label{omcrit}
\end{equation}

The two definitions (\ref{omsp}) and (\ref{omcrit}) yield in a sense
two extreme estimates of the hard pomeron singularity
($\om_c(t)\lesssim\op(t)\lesssim\om_{s}(t)$) of the full anomalous
dimension series. Both will be discussed here on the basis of our
improved BFKL kernel.

%========================= 3 =========================

\section{Improved subleading kernels}

The general form (\ref{seriet}) of the kernel of the RG improved small-$x$
equation is strongly constrained by ($i$) the exact leading and
next-to-leading $\log s$ calculations~\cite{bfkl,fl98,cc98} and ($ii$)
the collinear singularity structure of Eq.~(\ref{chin}). This leads to
a natural identification of the coefficient kernels $K_0^{\om}$ and
$K_1^{\om}$ --- up to some NNL ambiguity --- following the procedure
of Refs.~\cite{s98,cd99} which is described in detail here.

\subsection{Form of the collinear singularities}\label{ss:fcs}

Let us first recall the argument leading to Eq.~(\ref{chin}). The RG
invariant kernel in Eq.~(\ref{tetto}) acquires collinear singularities 
in the limit $k'/k\to0$ ($k/k'\to0$), which corresponds to strong
ordering of the transverse momenta in the direction of the ``upper''
scale $k^2$ (``lower'' scale $k_0^2$). Therefore, such singularities are
easily expressed for the kernel $\K^{(u)}$ ($\K^{(l)}$) corresponding
to energy-scale $k^2$ ($k_0^2$) in the NL $\kk$-factorization formula
(\ref{fatt}). For $k\gg k'$, $\K^{(u)}$ acquires the form
\begin{align}
 \K_{\om}^{(u)}(\kk,\kk')&\simeq{\ab(t)\over k^2}\;\exp\int_{t'}^t
 \tilde{\ga}(\om,\as(\tau))\;\dif\tau\qquad(t\gg t')\nonumber\\
 &={\ab(t)\over k^2}\left(1-b\ab(t)\log{k^2\over k'{}^2}\right)^
 {-{A_1(\om)\over b}}\,,\label{kupper}
\end{align}
where $\tilde{\ga}$ is the non-singular part of the gluon anomalous
dimension of Eq.~(\ref{dimanom}), the singular one being accounted for 
by the BFKL iteration itself.

Expanding Eq.~(\ref{kupper}) in $\ab(t)$ and comparing with the
general definition (\ref{seriet}), leads to the identification of the
kernels $K_n^{(u)\,\om}$ in the collinear limit, whose eigenvalue
functions turn out to have the singularities
\begin{equation}
 \chi^{(u)\,\om}_n(\ga)\simeq{1\cdot A_1(A_1+b)\cdots(A_1+(n-1)b)\over
 \ga^{n+1}}\,,\qquad(\ga\ll1)\,. \label{chiupper}
\end{equation}
which correspond to single logarithmic scaling violations for $k\gg
k_0$.  A similar reasoning yields the collinear behavior of
$\K^{(l)\,\om}$ in the opposite strong ordering region $k'\gg k$
\begin{align}
 K_{\om}^{(l)}(\kk,\kk')&\simeq{\ab(t')\over k'{}^2}\;\exp\int_t^{t'}
 \tilde{\ga}(\om,\as(\tau))\;\dif\tau\qquad(t'\gg t)\nonumber\\
 &={\ab(t)\over k'{}^2}\left(1-b\ab(t)\log{k'{}^2\over k^2}\right)^
 {{A_1(\om)\over b}-1}\label{klower}
\end{align}
and to the singularities
\begin{equation}
 \chi_n^{(l)\,\om}(\ga)\simeq{1\cdot(A_1-b)\cdots(A_1-nb)\over
 (1-\ga)^{n+1}}\,,\qquad(1-\ga\ll1)\,.\label{chilower}
\end{equation}

However, the similarity relation (\ref{cambioscala}) connects the
kernels $\K^{(u)}$ and $\K^{(l)}$. Therefore $\K^{(u)}$ has the
singularities (\ref{chilower}) shifted at $\ga=1+\om$ also, and
similarly $\K^{(l)}$ has the singularities (\ref{chiupper}) shifted at 
$\ga=-\om$. As a consequence, the symmetrical kernel $\K_{\om}$ --- for 
the energy-scale $s_0=kk_0$ --- has both kinds of singularities shifted 
by $\pm\om/2$, as anticipated in Eq.~(\ref{chin}). In particular the
leading and NL coefficient kernels have singularities
\begin{align}
 \chi_0^{\om}(\ga)&\sim{1\over\ga+\ts{1\over2}\om}+
 {1\over1-\ga+\ts{1\over2}\om}\\
 \chi_1^{\om}(\ga)&\sim{A_1(\om)\over(\ga+\ts{1\over2}\om)^2}+
 {A_1(\om)-b\over(1-\ga+\ts{1\over2}\om)^2}\,.
\end{align}

Note the $b$-dependent asymmetry of the singularities in
Eq.~(\ref{chin}) under the $\ga\leftrightarrow1-\ga$
transformation. It is due to the fact that the expansion
(\ref{seriet}) involves $\ab(t)$ (and not $\ab(t')$). Of course, the
kernel $\K_{\om}$ itself must be symmetrical under $t\leftrightarrow
t'$ exchange, so that expressing $\ab(t')$ in terms of $\ab(t)$
\begin{equation}
 \ab(t')={\ab(t)\over1+b\ab(t)\log\ds{k'{}^2\over k^2}}
\end{equation}
leads to the symmetry constraints
\begin{equation}
 \chi_n^{\om}(\ga)=\sum_{m\le n}
 {n\choose m}     %  {m\brace n} , {m\brack n}
 (-b\de_{\ga})^{n-m}\chi_m^{\om}(1-\ga)\,,   \label{vincolo}
\end{equation}
where $\de_{\ga}$ denotes the $\ga$-derivative. It is straightforward
to check by the binomial identity
\begin{equation}
 {r+n\choose n}=\sum_{m=0}^n{r\choose m}{n\choose m}
\end{equation}
that the symmetry constraints (\ref{vincolo}) are indeed satisfied by
Eq.~(\ref{chin}). In particular we must have
\begin{equation}
 \chi_0^{\om}(1-\ga)=\chi_0^{\om}(\ga)\,,\qquad
 \chi_1^{\om}(1-\ga)=\chi_1^{\om}(\ga)+b \chi_0^{\om}{}'(\ga)\,,
 \label{simmetria}
\end{equation}
showing that the antisymmetric part of $\chi_1^{\om}(\ga)$ is
$-{b\over2}\chi_0^{\om}{}'(\ga)$.

\subsection{Form of the leading coefficient kernel}

Given the fact that the $\om$-dependence is tied up with the
$\ga$-dependence in the singularities (\ref{chin}), it follows that
the leading $\log s$ hierarchy, corresponding to a pure
$\om$-expansion at fixed $\ga$, is poorly convergent close to $\ga=0$
and $\ga=1$. This observation follows from the trivial expansion
\begin{equation}
 {1\over\ga+\ts{1\over2}\om}={1\over\ga}\left(1-{\om\over2\ga}+
 \left({\om\over2\ga}\right)^2+\cdots\right)    \label{espan}
\end{equation}
and was used in Ref.~\cite{s98} to suggest a resummed form of the
leading kernel eigenvalue function
\begin{align}
 \chi^{\om}_0(\ga)=&\left[\psi(1)-\psi(\ga+\ts{1\over2}\om)\right]+
 \left[\psi(1)-\psi(1-\ga+\ts{1\over2}\om)\right]\label{lund}\\
 =&\chi_0(\ga)-{1\over2}\om{\pi^2\over\sin^2\pi\ga}+\cdots
 \,.\nonumber
\end{align}
The kernel $K_0^{\om}$, corresponding to Eq.~(\ref{lund}) is that
occurring in the Lund model~\cite{ags96} and is given by
\begin{equation}
 K_0^{\om}(\kk,\kk')=K_0(\kk,\kk')\left({k_<\over k_>}\right)^{\om}
 \,,        \label{soglia}
\end{equation}
where $k_>\equiv\max(k,k')$ and $k_<\equiv\min(k,k')$.
It is thus related to the customary leading kernel $K_0$ by the
``threshold factor'' $(k_</k_>)^{\om}$. This means that the
$s$-dependence provided by its inverse Mellin transform is
\begin{equation}
 K_0(s;\kk,\kk')\equiv\int{\dif\om\over2\pi\ui}\left(
 {s\over kk'}\right)^{\om}K_0^\om(\kk,\kk')=K_0(\kk,\kk')\,
 \Th(s-k_>^2)\,. \label{sogliola}
\end{equation}

Can one justify the form of the kernel (\ref{soglia}) ``a priori''?
From the point of view of the RG improved equation, any kernel which
($i$) reduces to $K_0$ in the $\om\to0$ limit and ($ii$) has
the leading simple poles of Eq.~(\ref{chin}) for $n=0$, is an
acceptable starting point. An alternative choice of this kind will
differ from $K_0^{\om}$ by a NL kernel {\em without} $\ga=0$ or
$\ga=1$ singularities. The resulting ambiguity can thus be reabsorbed
by a proper subtraction in the NL coefficient kernel.

Nevertheless, the threshold interpretation of Eqs.~(\ref{soglia}) and
(\ref{sogliola}) is appealing. For instance, the first iteration of
such a kernel provides the expression
\begin{align}
 K_0^2(s;\ku,\kd)=&\int{\dif\om\over2\pi\ui}\left(
 {s\over k_1k_2}\right)^{\om}\left({1\over\om}K_0^{\om}\right)^2
 \label{itero}\\
 =&\int{\dif^2\kk\over\pi}\;K_0(\ku,\kk)\left(\log{s\over k_1k_2}-
 \eta(k_1,k)-\eta(k_2,k)\right)K_0(\kk,\kd)\nonumber
\end{align}
where
\begin{equation}
 \cosh\eta(k_i,k)\equiv{k^2+k_i^2\over2kk_i}\,. \label{eta}
\end{equation}
The threshold condition implied by Eq.~(\ref{itero})
\begin{equation}
 {s\over2k_1k_2}=\cosh\eta>\cosh(\eta(k_1,k)+\eta(k,k_2)) \label{toller}
\end{equation}
is reminiscent~\cite{c98b} of phase space in Toller variables~\cite{t65}
and may be regarded as an alternative way of stating coherence
effects~\cite{c88,cfm90}, as implied in the original version of the
Lund model itself~\cite{ags96}.

Whether or not such hints will eventually provide a more direct
justification of $K_0^{\om}$, the fact remains that Eq.~(\ref{lund})
resums the $\om$-dependence of the $\ga$-singularities, and thus
provides the correct singularities of the scale-dependent terms of the 
NL kernel. Therefore, it is a good starting point, yielding NL
contributions which are smoother than those in the $\as(t)$-expansion, 
as we now discuss.

\subsection{Form of the next-to-leading contribution}\label{ss:fnlc}

The NL contribution $K_1^{\om}$ is constructed by requiring that ($i$)
the Green's function $\G_{\om}$ reproduce the known NL calculations
and ($ii$) the collinear singularities be as in Eq.~(\ref{chin}) with
$n=1$.

In order to implement condition ($i$) we have first to relate the
$\om$-dependent formulation of $\G_{\om}$ in Eq.(\ref{defg}) to the
customary expression of the BFKL kernel at NL level
\begin{equation}
 {1\over\om}\K={1\over\om}\left(\ab K_0+\ab^2K_1+\cdots\right)
 \,.     \label{ksuom}
\end{equation}
The $\om$-dependent formulation of Eq.~(\ref{seriet}) yields instead
the NL expansion
\begin{align}
 {1\over\om}\K^{\om}=&{1\over\om}\left(\ab K_0^{(0)}+\ab\om K_0^{(1)}
 +\ab^2K_1^{(0)}+\cdots\right)\,,    \label{komsuom}\\
 &K_i^{\om}\equiv K_i^{(0)}+\om K_i^{(1)}+\cdots\,, \nonumber
\end{align}
which is actually more general than Eq.~(\ref{ksuom}) because the
$\ab\om$ term, coming from the $\om$-expansion of $K_0^{\om}$, is a
possible NL contribution too.

Now it turns out that, at NL level, the formulation (\ref{komsuom})
reduces to the one in (\ref{ksuom}), provided the impact-factor
kernels $H$, $H^{\dagger}$ of Eq.~(\ref{fatt}) are taken into
account. In fact, by using the expansion (\ref{komsuom}) and simple
operator identities, we can write
\begin{equation}\!\!
 \left(1-{1\over\om}\K^{\om}\right)^{-1}=\left(1-\ab K_0^{(1)}
 \right)^{-{1\over2}}\left(1-{1\over\om}(\ab K_0+\ab^2K_1+\cdots)
 \right)^{-1}\left(1-\ab K_0^{(1)}\right)^{-{1\over2}} \label{operatori}
\end{equation}
provided we set
\begin{equation}
 K_0=K_0^{(0)}\,,\qquad K_1=K_1^{(0)}+{1\over2}\left(K_0^{(1)}K_0+
 K_0K_0^{(1)}\right)\,. \label{identifico}
\end{equation}
Eqs.~(\ref{operatori}) and (\ref{fatt}) show that the two formulations
above differ by just a redefinition of the impact-factor kernels,
while Eq.~(\ref{identifico}) means that $K_1^{(0)}$ is given by $K_1$,
after subtraction of the term already accounted for in the
$\om$-dependence of $K_0^{\om}$. Using Eq.~(\ref{lund}) this yields
the $\om=0$ limit of the eigenvalue function
\begin{equation}
 \chi_1^{\om=0}(\ga)=\chi_1(\ga)+{1\over2}\chi_0(\ga)
 {\pi^2\over\sin^2\pi\ga}\,.  \label{chiom0}
\end{equation}

The subtraction term so obtained is important because it has cubic
poles at $\ga=0,1$ which cancel the corresponding ones occurring in
the energy-scale dependent part ($-{1\over4}\chi''_0$) of $\chi_1(\ga)$
found by Camici and one of us~\cite{cc98}, as noticed by another one
of us~\cite{s98} and seen explicitly in Eq.~(\ref{chi1}). Furthermore,
the impact-factor kernels of Eq.~(\ref{operatori}) have quadratic
poles which similarly account for the ones occurring in $H$ and
$H^{\dagger}$~\cite{cc98,cd98}. This means that the remaining
contributions are, in both cases, much smoother in the $\om$-dependent
formulation.

In order to implement condition ($ii$) on $\chi_1^{\om}$, we note that
the $\om=0$ limit (\ref{chiom0}) still contains double and single
poles at $\ga=0,1$, which should be shifted according to
Eq.~(\ref{chin}). By neglecting the (small) $q\bar{q}$ contributions,
the explicit form of Eq.~(\ref{chiom0}), following from
Refs~\cite{fl98,cc98} for the energy-scale $s_0=kk_0$, is
\begin{align}
 \chi_1^{\om=0}(\ga)=&-{1\over2}\left({11\over12}(\chi_0^2(\ga)+
 \chi_0'(\ga))\right)+\left[-{1\over4}\,\chi_0''(\ga)+{1\over2}
 \chi_0(\ga){\pi^2\over\sin^2\pi\ga}\right]+\nonumber\\
 -&{1\over4}\left\{\!
 \left({\pi\over\sin\pi\ga}\right)^2\!\!{\cos\pi\ga\over3(1-2\ga)}\left(11+
 {\ga(1-\ga)\over(1+2\ga)(3-2\ga)}\right)\right\}+\nonumber\\
 +&\left({67\over36}-{\pi^2\over12}\right)\!
 \chi_0(\ga)+{3\over2}\,\zeta(3)+{\pi^3\over4\sin\pi\ga}-\Phi(\ga)  
 \,,\label{chi1}\\
 \Phi(\ga)\equiv&\sum_{n=0}^{\infty}(-)^n\left[{\psi(n+1+\ga)-
 \psi(1)\over(n+\ga)^2}+{\psi(n+2-\ga)-\psi(1)\over(n+1-\ga)^2}\right]
 \,.\nonumber
\end{align}
Here we have singled out some singular terms which have a natural
physical interpretation, namely the running coupling terms (in round
brackets), the energy-scale-dependent terms (in square brackets) and
the collinear terms (in curly brackets).

The running coupling terms have a double pole at $\ga=1$ only, and
account for the asymmetric part of $\chi_1$ (given by
$-{b\over2}\chi'_0$) 
which provides the $b$-dependent double pole on $\chi_1^{\om}$ in
Eq.~(\ref{chin}). The collinear terms have symmetrical double poles
with residue $A_1(\om=0)$, in accordance with Eq.~(\ref{chin})
also. Both types of singularities can be shifted by adding a NNL term, 
vanishing in the $\om=0$ limit, which we take to be
\begin{equation}
 A_1(\om)\psi'(\ga+\ts{\om\over2})-A_1(0)\psi'(\ga)+(A_1(\om)-b)
 \psi'(1-\ga+\ts{\om\over2})-(A_1(0)-b)\psi'(1-\ga)\;.   \label{shift}
\end{equation}
This term incorporates the $\om$- dependence of the one-loop anomalous
dimension (\ref{dimanom}) too.

The energy-scale-dependent term in square brackets contains the
subtraction (\ref{chiom0}) and has, therefore, simple poles at
$\ga=0,1$ only, which we can shift by adding the contribution
\begin{equation}
 {\pi^2\over6}\left(\chi_0^{\om}(\ga)-\chi_0(\ga)\right)\,.\label{add}
\end{equation}
By then collecting Eqs.~(\ref{chiom0}), (\ref{shift}) and (\ref{add})
we obtain the final eigenvalue function
\begin{equation}
 \chi_1^{\om}(\ga)\equiv\tilde{\chi}_1(\ga)+A_1(\om)\psi'(\ga+\ts{1\over2}
 \om)+(A_1(\om)-b)\psi'(1-\ga+\ts{1\over2}\om)+\ds{\pi^2\over6}
 \chi_0^{\om}(\ga)\,,     \label{chi1om}
\end{equation}
where
\begin{equation}
 \tilde{\chi}_1(\ga)\equiv\chi_1(\ga)+{1\over2}\chi_0(\ga)
 {\pi^2\over\sin^2\pi\ga}-{\pi^2\over6}\chi_0(\ga)-A_1(0)\psi'(\ga)
 -(A_1(0)-b)\psi'(1-\ga)      \label{chitilde}
\end{equation}
is a symmetrical function without $\ga=0$ or $\ga=1$ singularities at
all.
The expression (\ref{chi1om}) satisfies in addition the symmetry
constraints (\ref{simmetria}), having antisymmetric part
$-{b\over2}\chi_0^{\om}{}'$.

Of course, there is some ambiguity involved in the choice of the
subtraction terms (\ref{shift}, \ref{add}), which boils down to the
possibility of adding to (\ref{chitilde}) a term, vanishing in the 
$\om=0$ limit, and having only higher twist $\ga$-singularities,
around $\ga=-1,-2,\cdots$ and $\ga=2,3,\cdots$. This ambiguity leads
to an error which is of the same order as that made in the NL
truncation of the $\om$-expansion of the solution in Sec.~\ref{s-oe},
as we shall see next.

\subsection{Numerical importance of collinear effects at NLO}
\label{sec:collval}

\begin{figure}[htbp]
  \begin{center}
    \epsfig{file=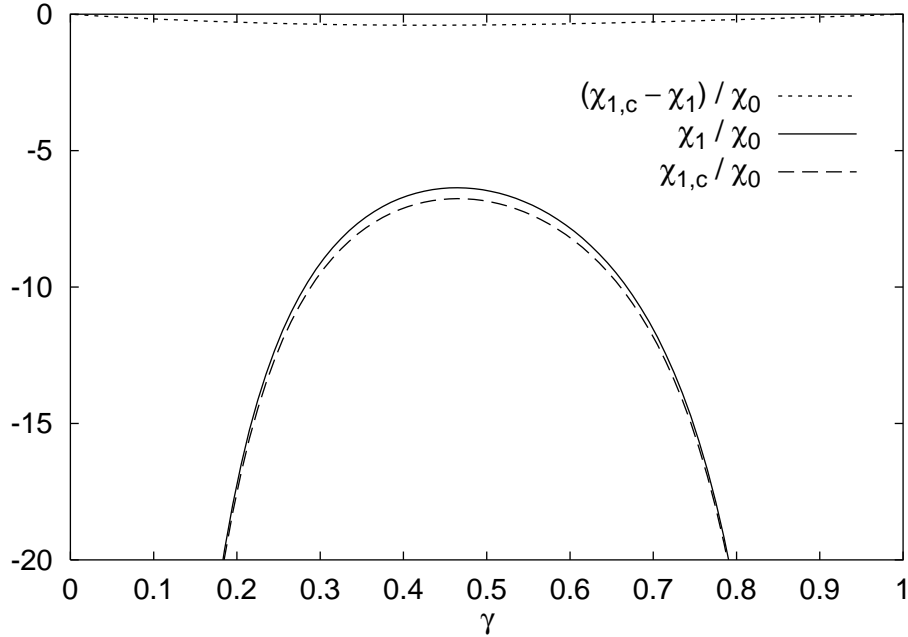}
    \caption{A comparison of the collinearly-enhanced (double and
      triple poles only) part of the NLO corrections with the full
      NLO corrections; $\nf=0$.}
    \label{fig:chi1c}
  \end{center}
\end{figure}

Above we have given the general form for the collinear singularities
of the kernel at all orders. It is of interest to consider at NLO just
how much of the full corrections come from these collinearly enhanced
terms. Accordingly we look at the part of the NLO corrections which
contains just double and triple poles, $\chi_{1,c}$:
\begin{equation}
  \chi_{1,c} = \frac{A_1}{\ga^2}
                + \frac{A_1 - b}{(1-\ga)^2}
                - \frac{1}{2\ga^3}
                - \frac{1}{2(1-\ga)^3}\,.
\end{equation}
This is compared with the full $\chi_1$ in figure~\ref{fig:chi1c},
where we have plotted their ratios to $\chi_0$. The remarkable
observation is that over a range of $\ga$, the collinear approximation
reproduces the true corrections to within $7\%$.  It is obviously
impossible to say whether this is true at higher orders as well.
However the fact that the study of collinear terms has such predictive
power at NLO is a non-trivial point in favour of our resummation
approach.

%========================= 4 ========================

\section{Improved next-to-leading solution}

Having constructed the coefficient kernels $K_0^{\om}$ and $K_1^{\om}$
with consistent collinear behavior (cf.\ Eqs.~(\ref{lund}) and
(\ref{chi1om})) we would like to know the large-$t$ behavior of the
solutions of the improved small-$x$ equation, whose kernel (\ref{seriet})
is truncated at NL level. This problem has been solved in general in
Sec.~2, and we describe here the NL features.

\subsection{$\om$-Expansion of the gluon distribution}

According to Sec.~2.4, the eigenfunctions $\F_{\om}^{\mu}(\kk)$
($-\infty<\mu<\mp(\om)$) can be found in the small-$\mu$, large-$t$
regime
\begin{equation}
 bt\gtrsim{1\over\mu}\gtrsim{1\over\om}\gg1   \label{regime}
\end{equation}
by the $\ga$-representation (\ref{rappres}), i.e.,
\begin{subequations}
\begin{equation}
 k^2\F_{\om}^{\mu}(\kk)=\int\camint{\dif\ga\over2\pi\ui}\exp\left\{\ga t-
 {1\over b\mu}X_\om(\ga,\mu)\right\}\,,   \label{rap}
\end{equation}
where the exponent function $X_\om$ is provided by the small-$\mu$
expansion
\begin{equation}
 \de_{\ga}X_\om(\ga,\mu)\equiv\chi_{\om}(\ga,\mu)=
 \chi_0^{\om}(\ga)+\mu\,{\chi_1^{\om}(\ga)\over\chi_0^{\om}(\ga)}+
 \mu^2\,\eta_2^{\om}(\ga)+
 \mu^3\,\eta_3^{\om}(\ga)+\cdots      \label{muesp}
\end{equation}\label{rapfin}
\end{subequations}
and $\eta_2^{\om}$, $\eta_3^{\om}$, \dots are given in
Eq.~(\ref{ete}).

Furthermore, by the factorization property (\ref{fattg}), valid for
$|t-t_0|\gg1$, the gluon Green's function (\ref{g}) is itself
proportional to $\F_{\om}(\kk)\equiv\F_{\om}^{\mu=\om}(\kk)$, which is
obtained by setting $\mu=\om$ in Eqs.~(\ref{rapfin}), i.e.,
\begin{equation}
 \de_{\ga}X_\om(\ga,\om)\equiv\chi(\ga,\om)=
 \chi_0^{\om}(\ga)+\om\,{\chi_1^{\om}(\ga)\over\chi_0^{\om}(\ga)}
 + {\rm NNL}\,,    \label{omesp}
\end{equation}
where we have now truncated the expansion to NL level. The ensuing
error is argued to be small (Sec.~\ref{ss:eoe}).
The RG regime holds if there is a stable saddle point
\begin{equation}
 b\om t=\chi(\ga,\om)\simeq\chi_0^{\om}(\ga)+\om\,{\chi_1^{\om}(\ga)
 \over\chi_0^{\om}(\ga)}+{\rm NNL}\,,\qquad\chi'(\ga,\om)<0\,,
 \label{ps}
\end{equation}
which dominates the large-$t$ behavior of Eq.~(\ref{rap}), providing
the anomalous dimension representation (\ref{gdiman}). The effective
anomalous dimension can be continued past its saddle point value by
means of Eqs.~(\ref{derlog}) and (\ref{gint}), which use the
$\ga$-representation (\ref{rap}) for $\mu=\om$.

\subsection{Properties of the kernel and its solutions}\label{sec:kernprop}

In this section we illustrate some of the features of the resummed
kernel and of the regular solution as obtained with the
$\ga$-representation.

\begin{figure}[htbp]
  \begin{center}
    \epsfig{file=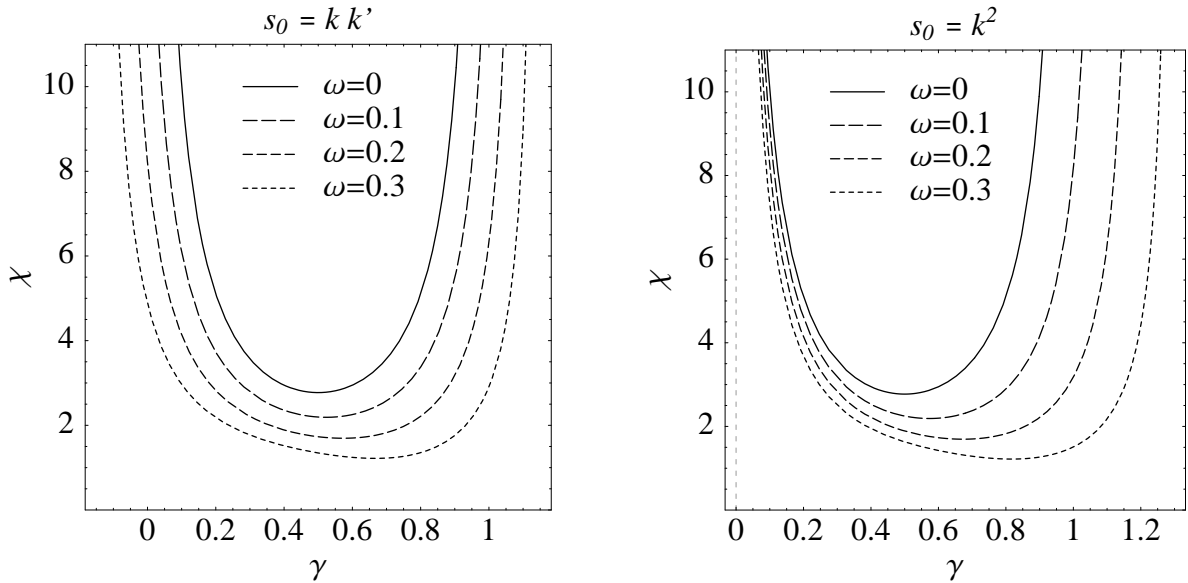}
    \caption{$\chi(\ga,\om)$ as a function of $\ga$ for various values
                 of $\om$, for the symmetric energy-scale $s_0=kk'$ on 
                 the left and for $s_0=k^2$ on the right. Here $\nf=0$.}
    \label{fig:chigaom}
  \end{center}
\end{figure}

It can be instructive to examine the resummed eigenvalue function
$\chi(\ga,\om)$ in Eq.~(\ref{omesp}) in two different ways. Firstly as
a function of $\ga$ for various values of $\om$, as shown in
Fig.~\ref{fig:chigaom}.
In the symmetric energy-scale case, we can see how the original
$\ga$-poles at $\ga=0,1$ are displaced in a symmetric way for
$\om\neq0$. The slight asymmetry of $\chi(\ga,\om)$ is due to
Eq.~(\ref{simmetria}), which in turn comes from the perturbative
expansion in $\ab(t)$, as discussed in Sec.~\ref{ss:fcs}. Even for
sizeable values of $\om$ the eigenvalue function preserves its shape,
with a stable minimum ($\chi''(\ga_m,\om)>0$) around $\ga=1/2$. This
stability is a necessary condition to avoid the oscillating behavior
noticed in Ref.~\cite{r98}.

With the ``upper'' scale choice $s_0=k^2$, the pole at $\ga=0$ is
$\om$-independent as it should, while the pole at $\ga=1$ is shifted
for $\om\neq0$. In this case we also have a stable minimum, in a
slightly different position with respect to the previous case
($\ga_m\to\ga_m+\ho$).

\begin{figure}[htbp]
  \begin{center}
    \epsfig{file=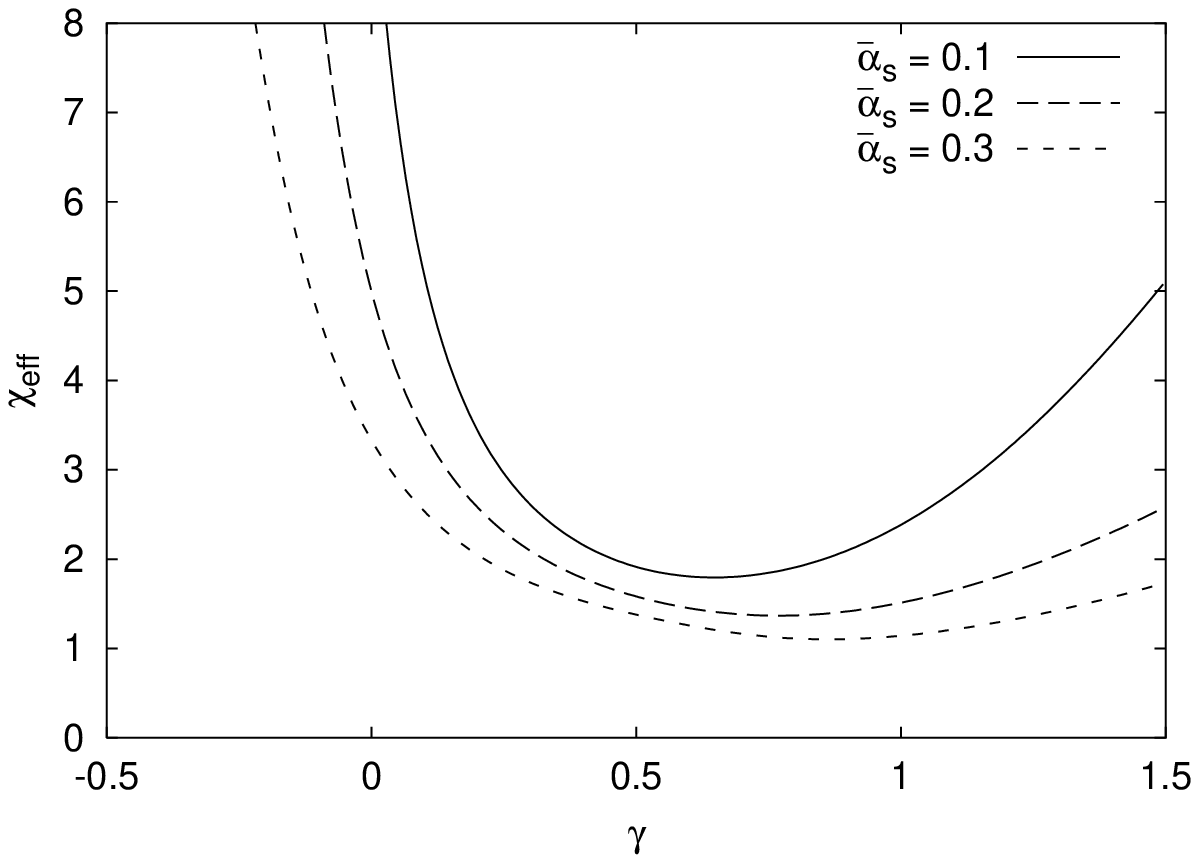}
    \caption{$\chieu(\ga,\ab)$ as a function of $\ga$ for various
            values of $\ab$, for energy-scale $s_0=k^2$ and $\nf=0$.} 
    \label{fig:chieff}
  \end{center}
\end{figure}

A second way of looking at the resummed kernel is to examine a
quantity which we call $\chieu(\ga,\ab)$, defined by
\begin{equation}
 \chieu(\ga,\ab) = \chi^{(u)}(\ga,\om=\ab\chieu)\,.
  \label{chieff}
\end{equation}
This is closely related to the saddle-point approximation for
evaluating the $\ga$-re\-pre\-sen\-ta\-tion, \eqref{gint}, since the
value of the saddle-point, $\gb$ satisfies $\om=\ab\chieu(\gb,\asb)$;
$\gb$ in such a case is itself closely related to the effective
anomalous dimension, \eqref{gaeff}. In Fig.~\ref{fig:chieff} we show
$\chieu(\ga,\ab)$ for different values of $\ab$.  The marked asymmetry
is due to the energy-scale choice $s_0=k^2$.  We note the rather
different structure from the $\chi(\ga,\om)$ shown in
Fig.~\ref{fig:chigaom}. In particular there are no longer any
divergences. That on the right is shifted by an amount $\om$: as a
result rather than a pole one has $\chieu\sim\ga/\ab$, as discussed in
\cite{s98}. That on the left instead becomes
$\chieu\sim\ab\esp{-\ga/\ab}$ for negative $\ga$ as a result of the
inclusion of the dependence on the DGLAP splitting function (in
particular the $1/(1-z)$ part, which gives $A_1(\om)\simeq-\log\om$
for $\om\to+\infty$).  Another feature of $\chieu$ worth noting (though
not immediately visible from Fig.~\ref{fig:chieff}) is that for
$\nf=0$, $\ab\chieu(0,\asb) = 1$, independently of $\asb$. This is so
because close to $\ga=0$
\begin{equation}
  \om = \ab\chieu = \ab \frac{1 + \om A_1(\om)}{\ga} +
  \ord(\ab)\,. 
  \label{omga0}
\end{equation}
For $\ga=0$, we have that $1 + \om A_1(\om) = 0$. Since $A_1(1)=-1$,
$\om=1$ at $\ga=0$.  This point is of significance insofar as it
relates to the problem of ensuring conservation of the momentum
sum-rule for the gluon distribution (its $\om=1$ moment).
\begin{figure}[htbp]
  \begin{center}
    \epsfig{file=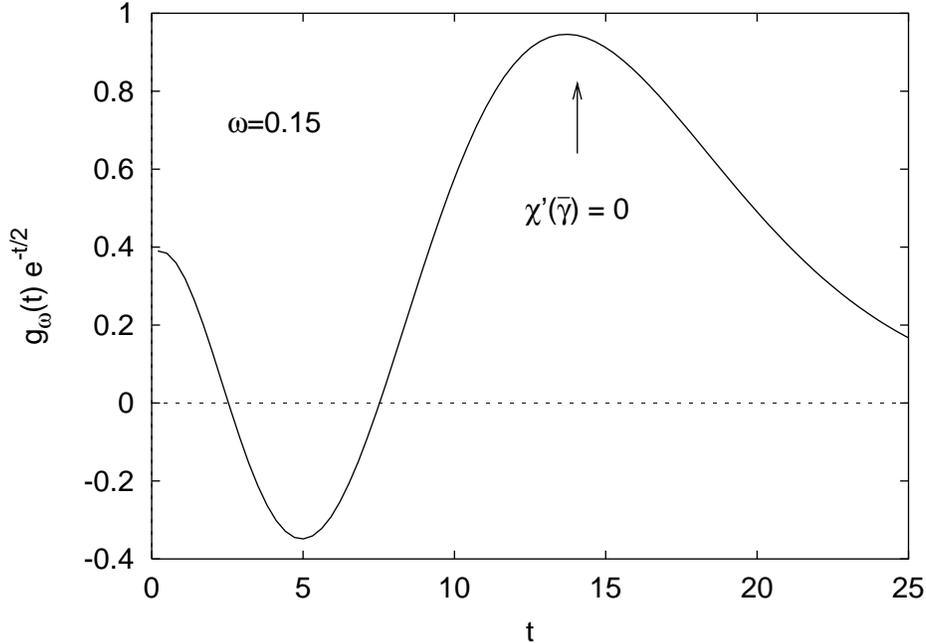}
    \caption{$g_\om(t) \esp{-t/2}$ for $\om=0.15$, with energy-scale
      $s_0=k^2$. The normalization is arbitrary.}
    \label{fig:kF}
  \end{center}
\end{figure}

Next we examine the properties of the regular solution of the
small-$x$ equation for the integrated gluon distribution in
Eq.~(\ref{gint}). In order to concentrate on the region most relevant
for a consideration of the small-$x$ properties of the anomalous
dimension, in Fig.~\ref{fig:kF} we actually show $g_\om(t)
\esp{-t/2}$, as a function of $t = \ln k^2/\Lambda^2$. There are two
critical points on the curve.  Firstly the point labelled
$\chi'(\gb)=0$, namely where the saddle-point solution $b\om t=
\chi(\gb,\om)$ sits at the minimum of $\chi$ for that $\om$. Let us
refer to that point as $t_s$.  The second point of interest, which we
call $t_c$, is the rightmost zero of $g_\om$.  This is the point where
the effective anomalous dimension has a divergence.  Since the
solution in this region has roughly the form~\cite{l86}
\begin{equation}
  g_\om(t) \sim \mathrm{Ai}\left(\left(\frac{2}{\chi_m'' b^2
        \om^2}\right)^{1/3} (b\om t - \chi_m)\right).
  \label{gomAiry}
\end{equation}
One can estimate the difference between $t_c$ and $t_s$ as being
\begin{equation}
  t_s - t_c =\xi_0\left(\frac{\chi''_m}{2 b \om}\right)^{1/3} +
  \cO{1}\,,
  \label{tscdiff}
\end{equation}
where $-\xi_0\simeq-2.3381$ is the position of the rightmost zero of the
Airy function. For fixed $\as$, this translates to a difference
between the estimates for $\om_s$ and $\om_c$ which, for $\ab\ll1$,
has the following form:
\begin{equation}
  \om_s - \om_c =\xi_0\left(\frac{b^2 \chi_m^2
      \chi_m''}{2}\right)^{1/3} \asb^{5/3} + \cO{\asb^2} \simeq 11.16
  \,\asb^{5/3}\,. 
  \label{omscdiff}
\end{equation}
Such $\as^{5/3}$ contributions to $\om$ have already been observed in
other contexts where there is some form of cutoff on transverse
momenta, such as a running coupling which is zero below a certain
value of $t$, or non-forward elastic scattering (where the exchanged
transverse momentum places an effective cutoff on transverse momenta
in the evolution)~\cite{hr92,FRBook}.  Numerical estimates (based
on the $\gamma$-representation) for the difference between $\om_s$ and
$\om_c$ coincide with \eqref{omscdiff} but only for very small $\asb$.

For typical values of $\asb$, we note that $\om_s-\om_c \simeq 11.16
\asb^{5/3}$ is of the same order as the NLL corrections. It too, as
pointed out in~\cite{hr92,FRBook}, is the first term of a poorly
convergent series. The resummation procedure that we recommend (and
adopt) is to define $\om_c$ not through the power series in $\asb$,
but by looking for the rightmost zero of the regular solution.

\subsection{Estimate of error}\label{ss:eoe}

The question now arises: what is the {\em error} that we make in the
NL truncation of the RG improved equation? Our claim is that, in the
improved formulation, based on the $\om$-expansion (\ref{omesp}), this 
error is smaller than in the formal NL expansion in $\as(t)$. Let us
in fact estimate the remaining terms in Eq.~(\ref{muesp}). According
to Eq.~(\ref{chin}) further subleading eigenvalue functions contain at 
least higher order collinear poles which contribute to $\eta_2^{\om}$,
$\eta_3^{\om}$ and so on. A first observation is that, even if
$\chi_n^{\om}$ has ($n+1$)-th order poles, the $\eta_n^{\om}$'s have
at most {\em simple} poles, due to the powers of $\chi_0^{\om}$
in the denominator, roughly due to the replacement
$\ab(t)\sim\om/\chi_0^{\om}$. Therefore, their contribution cannot be
too big, even for small values of $\ga=\ord(\om)$.

Furthermore, one can check that, if $q\bar{q}$ contributions
(Sec.~4.2) are neglected, the leading collinear poles actually
{\em cancel out} in the expansions (\ref{ete}) of $\eta_2^{\om}$,
$\eta_3^{\om}$, \dots around both $\ga=0$ and $\ga=1$.
The mechanism of this cancellation can be cleared up as follows.

From the mathematical point of view , it is possible to have the
truncated NL solution to be an {\em exact} solution of
Eq.~(\ref{eqbfkl}), provided the following recurrence relations hold
(App.~\ref{a:dergamma})
\begin{equation}
 {\chi_2^{\om}\over\chi_0^{\om}}=\left({\chi_1^{\om}\over\chi_0^{\om}}
 -b\de_{\ga}\right){\chi_1^{\om}\over\chi_0^{\om}}\,,\qquad
 {\chi_3^{\om}\over\chi_0^{\om}}=\left({\chi_1^{\om}\over\chi_0^{\om}}
 -b\de_{\ga}\right){\chi_2^{\om}\over\chi_0^{\om}}\,,\qquad
 \cdots\,.           \label{ricorrenza}
\end{equation}
It is now really simple to check that such relations build up the
collinear singularities (\ref{chin}), which therefore must cancel out
in the subleading corrections $\eta_2^{\om}$, $\eta_3^{\om}$, \dots.
The recurrence relations (\ref{ricorrenza}) can also be interpreted as 
DGLAP equations in $\ga$-space, for the anomalous dimension
$\tilde{\ga}$ in Eq.~(\ref{dimanom}).

From a more physical point of view, it is not possible for simple
poles to survive in $\eta_2^{\om}$, $\eta_3^{\om}$, \dots because,
when replaced in the saddle point condition (\ref{ps}), they would
provide $\om^2$, $\om^3$, \dots corrections to the {\em one-loop}
anomalous dimensions which cannot possibly be there. In fact, the full
anomalous dimension is accounted for by Eqs.~(\ref{omesp},\ref{ps}) as
follows
\begin{equation}
 b\om t\simeq{1\over\gb}+{A_1(\om)\over\gb}\quad\Rightarrow\quad
 \gb=\ab\left({1\over\om}+A_1(\om)\right)\,,  \label{1loop}
\end{equation}
where we have taken the small-$\ga$ limit of the collinear safe
eigenvalue function $\chi^{(u)}(\ga,\om)$.

We therefore conclude that, in the purely gluonic case, the NL
$\om$-expansion (\ref{ps}) takes into account the collinear behavior
to all-orders, and that no further resummation is needed. This point
is perhaps more easily seen by replacing the NL truncation of
Eq.~(\ref{muesp}) in the saddle point condition (\ref{puntos}) to
yield the equation
\begin{equation}
 b\mu t=\chi_0^{\om}+\mu{\chi_1^{\om}\over\chi_0^{\om}}\quad
 \Rightarrow\quad
 \mu={\ab\chi_0^{\om}\over1-\ab\ds{{\chi_1^{\om}\over\chi_0^{\om}}}}
 \,.        \label{mueff}
\end{equation}
It is apparent from the last version of Eq.~(\ref{mueff}) that we are
dealing with an effective eigenvalue function which resums the
collinear behavior as a geometric series.

We are finally able to state that the error in the NL truncation
(\ref{muesp}) is uniformly $\ord(\om^2)$, the neglected coefficient
having {\em no $\ga=0$ nor $\ga=1$ singularities} at all. This error is
therefore of the same size as the ambiguity in the definition of
$\chi_1^{\om}$ that we have pointed out before. The corresponding
error in the saddle point condition (\ref{ps}) is a roughly
$\ga$-independent change of scale $\De(bt)=\ord(\om)$, or 
$\De(\as)=\ord(\om)\as^2$.

\subsection{Extension to $q\bar{q}$ contributions}\label{ss:qq}

The coefficient kernels $K_n^{\om}$ take up collinear singularities
not only from the nonsingular part of the gluon anomalous dimension
$\tilde{\ga}_{gg}$, but also from $q\bar{q}$ states which are coupled
to it in the one-loop gluon/quark-sea anomalous dimension matrix
\begin{equation}
 \tilde{\ga}_{ab}(\om)=\ab A_{ab}(\om)\equiv\ga_{ab}(\om)-\d_{ag}
 {\ab C_b\over N_c\om}\,,   \label{matrice}
\end{equation}
where $a=(q,g)$ and $C_a=(C_F,C_A)\equiv N_c(r,1)$ denote the partonic 
channels and color charges.

Although the numerical effect of quark-sea contributions to the gluon
anomalous dimensions is pretty small~\cite{cc97b}, including the
two-channel evolution (\ref{matrice}) changes the collinear problem
conceptually. While the small-$x$ equation stays of one-channel type, due
to the high-energy gluon exchange, the two-channel collinear behavior
yields two anomalous dimension eigenvalues
\begin{equation}
 \ga_{\pm}={\ga_{gg}+\ga_{qq}\over2}\pm\sqrt{\left({\ga_{gg}-
 \ga_{qq}\over2}\right)^2+\ga_{qg}\ga_{gq}}\,,     \label{gapm}
\end{equation}
with the approximate NL expansions
($\ds{\left[{\ga_{gq}/\ga_{gg}}\right]_{\rm Leading}=r}$)
\begin{equation}
 \ga_+\simeq\ga_{gg}+r\ga_{qg}\,,\qquad\ga_-\simeq\ga_{qq}-r
 \ga_{qg}\,,\qquad \left(r\equiv{C_F\over C_A}\right)\,.\label{gapiu}
\end{equation}

Recovering in the BFKL framework the full collinear behavior
(\ref{gapm}) is not trivial, because
$\ga_-\simeq-r\ga_{qg}^{(1)}=\ord(\as)$ starts at NL level and for
$\ga=\ord(\as)$ the leading $\log s$ hierarchy breaks down in the
$\as$-expansion~\cite{cc97b}. What do things look like in the
$\om$-expansion?

Note first that the derivation of the collinear behavior of
$K_n^{\om}$ in Sec.~3 can be repeated, by replacing $A_1(\om)$ with
the matrix $\A(\om)$ in Eq.~(\ref{matrice}), and by projecting the
final results onto the gluon channel, which corresponds to a bracket
between initial state ${0\choose1}$ and final state $(r\;1)$, because
the quark couples to the high-energy gluon with relative strength
$r=C_F/C_A$. Therefore, Eq.~(\ref{chin}) should be replaced by
\begin{align}
 \chi_n^{\om}(\ga)&\simeq{1\over(\ga+\ts{1\over2}\om)^{n+1}}
 \langle(r\;1)|\,1\cdot\A\cdot(\A+b)\cdots(\A+(n-1)b)\,|
 {0\choose1}\rangle\,,\qquad\ga\ll1\,,\label{chim}\\
 &\simeq{1\over(1-\ga+\ts{1\over2}\om)^{n+1}}
 \langle(r\;1)|\,1\cdot(\A-b)\cdots(\A-nb)\,|
 {0\choose1}\rangle\,,\qquad1-\ga\ll1\,.    \nonumber
\end{align}
In particular
\begin{align}
 \chi_1^{\om}&\simeq{\langle\A\rangle\over(\ga+\ts{1\over2}\om)^2}+
 {\langle\A\rangle-b\over(1-\ga+\ts{1\over2}\om)^2}\,,\label{chi1m}\\
 \chi_2^{\om}&\simeq{\langle\A(\A+b)\rangle\over
 (\ga+\ts{1\over2}\om)^3}+{\langle(\A-b)(\A-2b)\rangle\over
 (1-\ga+\ts{1\over2}\om)^3}\,,\nonumber
\end{align}
where $\langle\A\rangle=A_{gg}+rA_{qg}$, $\langle\A^2\rangle$,
\dots denote the brackets defined before in Eq.~(\ref{chim}).

Secondly, the kernel (\ref{chi1}) should be supplemented by the
($q\bar{q}$) contribution~\cite{cc96}, which completes the $b$-factor
in front of the running coupling terms and adds up a collinear
contribution, as follows
\begin{align}
 \chi_1^{q\bar{q}}(\ga)=&-{1\over2}\left({-2\nf\over12N_c}(\chi_0^2(\ga)+
 \chi_0'(\ga))\right)+\nonumber\\
 &-{\nf\over6N_c}\left\{{5\over3}\chi_0(\ga)+{3\over N_c^2}\,{\pi^2\over
 \sin^2\pi\ga}\,{\cos\pi\ga\over1-2\ga}\,{1+\ts{3\over2}\ga(1-\ga)\over
 (1+2\ga)(3-2\ga)}\right\}\,.     \label{chiqq}
\end{align}
Correspondingly, the subtraction term (\ref{shift}) changes by the
replacement\footnote{Since there is a (small) two-loop anomalous
dimension in the $Q_0$-scheme, induced by $q\bar{q}$ contributions,
one could envisage a shift of this simple pole in Eq.~(\ref{chiqq})
also, by a further change of the NL subtraction term.}
\begin{equation}
 A_{gg}=A_1\quad\to\quad\langle\A\rangle=A_{gg}+r
 A_{qg}\,,      \label{Agg}
\end{equation}
while $\chi_0^{\om}$ and the subtractions (\ref{chiom0}) and
(\ref{add}) are left unchanged.

The main differences with the purely gluonic case come out in the
$\om$-expansion of the solution, and specifically in the role of the
higher-order terms. In fact, if we repeat the calculation
(\ref{1loop}) with the new entries (\ref{chi1m}), we find
\begin{equation}
 \ga_+^{(1)}=\ab\left({1\over\om}+\langle\A\rangle\right)=\ab\left(
 {1\over\om}+A_{gg}(\om)+rA_{qg}(\om)\right)\,, \label{ga1piu}
\end{equation}
which is consistent with the NL expansion (\ref{gapiu}) for
$\ga_+$, but {\em is not} the full one-loop anomalous dimension
(\ref{gapm}).

Further terms in the $\om$-expansion must therefore contribute $1/\ga$ 
and $1/1-\ga$ poles, and they indeed do. From Eqs.~(\ref{gaeff}) and
(\ref{chi1m}) we find
\begin{equation}
 \De\ga_+^{(1)}=\left(\langle\A^2\rangle-\langle\A\rangle^2\right)\ab
 \om+\langle(\A-\langle\A\rangle)^3\rangle\ab\om^2+\cdots\label{dg}
\end{equation}
which checks with the explicit expansion of Eq.~(\ref{gapm}) up to the 
relevant order. The explicit matrix form of the corrections in
Eq.~(\ref{dg}) makes it clear why the two-channel problem allows the
survival of the simple $\ga$-poles at higher subleading orders.

Nevertheless, the small $\om$-expansion remains smoother than the
$\as$-expansion. In fact, the $\ord(\om^2)$ NNL terms being neglected
show simple poles only (around $\ga=0,1$), the general trend remains
the same as in Fig.~\ref{fig:chigaom}, provided $\om$ is not too
large. If $\om$ increases, $\ga_+$ decreases, and at some critical
value of $\om$, for which $\ga_+$ and $\ga_-$ become of the same
order, the $\om$-expansion will break down, eventually. Whether or not
the low-energy eigenvalue $\ga_-$ can still be described by an
all-order resummation in $\om$ remains an open question.

%========================= 5 ========================

\section{Anomalous dimension and hard pomeron}\label{s:adhp}

Here we present our main numerical results, for both the improved
gluon anomalous dimension and the hard pomeron, and we show their
stability.

\subsection{Results}\label{ss:res}

\begin{figure}[htbp]
  \begin{center}
    \epsfig{file=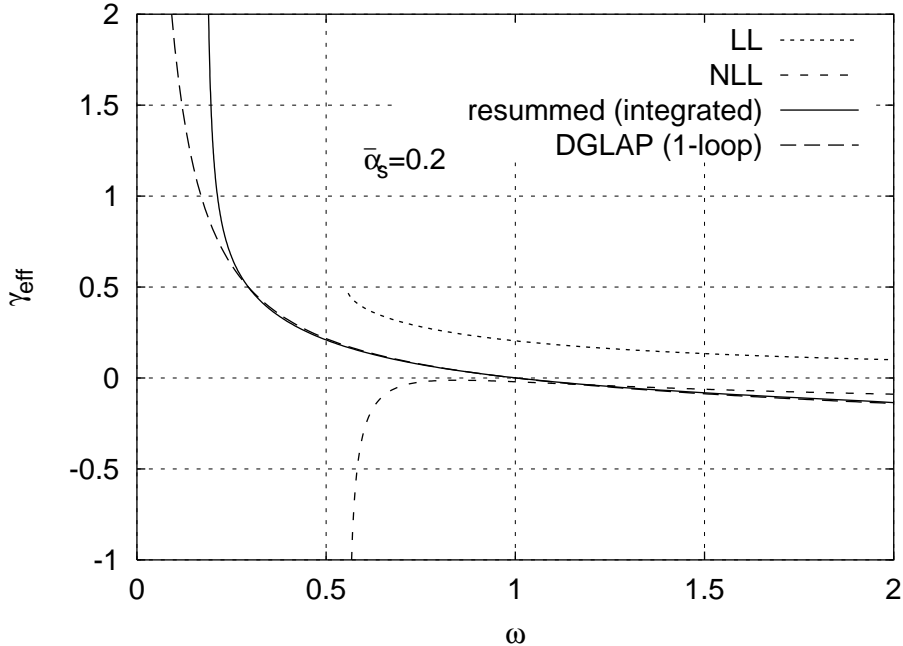}
    \caption{The anomalous dimension in various approximations}
    \label{fig:anomdim}
  \end{center}
\end{figure}
Fig.~\ref{fig:anomdim} shows the purely gluonic anomalous dimension as
a function of $\om$ for $\asb=0.2$. The LL anomalous dimension is just
$\ga_{LL}=\chi_0^{-1}(\om/\asb)$ and has the familiar branch-cut at
$\om=4\ln2 \asb$. The NLL anomalous dimension is taken as
\begin{equation}
  \label{NLLga}
  \ga_{NLL} =\ga_{LL} - \asb
  \frac{\chi_1(\ga_{LL})}{\chi_0'(\ga_{LL})}\,,
\end{equation}
and has the feature that it is always negative, with a divergent
structure around the same point as $\ga_{LL}$.  The resummed result,
defined in Eqs.~(\ref{gint}) and (\ref{derlog}), shows a divergence at
a much lower $\om$, defined by $\om_c(t)$ in Eq.~(\ref{omcrit}). What
is particularly remarkable is the similarity to the DGLAP result until
very close to the divergence. The momentum sum rule is automatically
conserved: for $\om=1$ we have $\gae=0$ (this is closely connected
with the fact that $\asb\chieu(0) = 1$) --- in past approaches the
need to impose this property in some arbitrary way was a major source
of uncertainty~\cite{ehw95,bf96,bv98}.

Another interesting feature of the resummed anomalous dimension is
that, for small $\ab$, the divergence at $\om_c$ is proportional to
$\asb^2$ and not to $\ab$:
\begin{subequations}\label{gae}
\begin{align}
  \gae &= - \frac{1}{\om-\om_c} \frac{d\om_c}{dt} +\cO{1}\\
  &\simeq\left(\chi_m-\frac53\left(\frac{b^2\chi_m^2{\chi_m}''}{2}
    \right)^{1/3}\xi_0\ab^{2/3} \right) \frac{b\,\asb^2}{\om-\om_c}
  \,,\qquad\qquad\om-\om_c \ll 1\,, 
\end{align}
\end{subequations}
which follows from the linear behavior of the regular solution close
to the zero, e.g., in the Airy representation of
Eq.~(\ref{gomAiry}). The singularity (\ref{gae}) causes the effective
splitting function to be power behaved for $x\to0$, i.e., $P_{{\rm
eff}}(x,t)\sim x^{-\om_c(t)}$.  Note however that, since $\om_c(t)$
comes from a zero of $g_{\om}(t)$, the singularity~(\ref{gae}) does
not necessarily transfer to $g_{\om}(t)$ itself which, according to
Eq.~(\ref{gint}), is expected to have an essential singularity at
$\om=0$ only, even if a complete analysis of possible singularities in
the complex $\om$-plane is still needed.
\begin{figure}[htbp]
  \begin{center}
    \epsfig{file=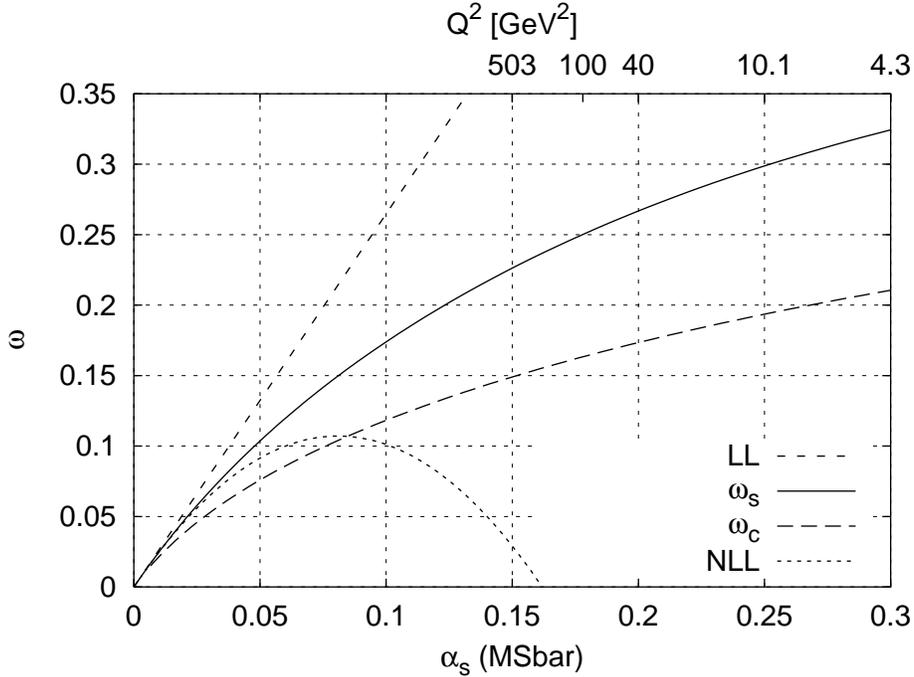}
    \caption{$\om_c$ and $\om_s$ as a function of $\as$ for
      the BFKL kernel with $\nf=0$.}
    \label{fig:omcoms}
  \end{center}
\end{figure}

The values of the exponents $\om_c$ and $\om_s$ as a function of $\as$
(and $Q^2$), are shown in Fig.~\ref{fig:omcoms} and compared with the
LL and pure NLL results. It is apparent that the improved equation
provides sensible predictions even for sizeable values of $\as$. A
significant difference between the two resummed exponents $\om_c$ and
$\om_s$ persists even to low values of $\as$, largely as a consequence
of their differing by a slowly convergent series of non-integer powers
of $\as$, as discussed in section~\ref{sec:kernprop}.

The above difference should not be too confusing. The exponent
$\om_s(t)$ signals the breakdown of the formal small-$x$ expansion of
the anomalous dimension of Eq.~(\ref{gaeff}), due to infinite
saddle-point fluctuations, while $\om_c(t)$ tells us the position of
the singularity of the resummed anomalous dimension. Their difference
arises from their different definitions, not from some instability of
our approach (cf.\ Sec.~\ref{ss:stab}).

What is the relation that such quantities bear to the pomeron
singularity $\op$, the leading $\om$-plane singularity of the gluon
Green's function? Though the latter is dependent on the strong
coupling region, we expect that $\op\gtrsim\underset{t}{\max}\om_c(t)$
for a positive definite $\as(t)$, due to the very definition of
$\om_c(t)$ as a zero of the integrated regular solution $g_{\om}(t)$,
to which $\F_{\om}(\kk)$ is closely related
(Sec.~\ref{sec:kernprop}). In fact $\op$ is defined as the value of
$\om$ being itself equal to the endpoint of the spectrum:
$\op=\mp(\op)$ (Sec.~\ref{ss:fots}), and thus corresponds to a
nodeless $\F_{\om}(\kk)$, regular for $t\to-\infty$ also. Therefore,
if the interaction does not change sign ($\as(t)>0$), $\F_{\om}(\kk)$
can have a node for $\om<\op$ only, so that $\om_c(t)<\op$.

The above remark implies that the small-$x$ behavior of the gluon
Green's function, dominated by the singularity at $\om=\op$ in
$\tilde{\F}_{\om}(\kk_0)$, is not sensitive to the region
$\om\simeq\om_c(t)$ where $\F_{\om}(\kk)$ changes sign. This fact is
consistent with the positivity constraint on the total cross section.

Furthermore, we can state that the frozen $\as$ regularization of
Sec.~\ref{ss:fots} maximizes the interaction strength in the strong
coupling region $t\lesssim\tb$, compared to various cut-off
procedures. Therefore we also expect $\op<\om_s(\tb)$, the value
quoted in Eq.~(\ref{omegap}) at the freezing point. It follows that
$\underset{t}{\max}\om_c(t)\lesssim\op\lesssim\underset{t}{\max}\om_s(t)$ 
or, in other words, that the two exponents of Fig.~\ref{fig:omcoms}
provide, in the strong coupling region, lower and upper bounds on the
pomeron intercept $\op$. Of course, the precise value of the latter
will be dependent on the size and shape of the effective coupling in
the small-$k$ region.

\subsection{Stability}\label{ss:stab}

The original $\mbox{LL}+\mbox{NLL}$ formalism suffered from
considerable instabilities under renormalization group scale and
scheme changes.

An important characteristic of any resummed approach is that it should
be relatively insensitive to such changes, and generally stable. 
In the approach advocated here, it has already been shown in the
previous sections that the formal truncation error is small. It 
still remains to demonstrate its stability in practice.

\paragraph{Renormalization scale and scheme.} 
Note first that in our approach the renormalization scale only enters
through the RG invariant $\La$ parameter (Eqs.~(\ref{seriet}) and
(\ref{gint})). It is then easy to see that the physical results are
$\La$-independent. A redefinition of $\La$ is essentially a shift in
$t$, say by an amount $\De t$.  There is a corresponding modification
of $\chi_1^{\om}$, $\chi_2^{\om}$, \dots by the amounts
\begin{equation}
  \chi_1^\om \to \chi_1^\om + b \De t \chi_0^\om\,,\qquad
 \chi_2^\om\to\chi_2^\om+2b\De t\chi_1^\om+b^2(\De t)^2
 \chi_0^\om\,,\qquad\dots      \label{deltachi}
\end{equation}
In the off-shell $\gamma$-representation (\ref{rapfin}), this corresponds
to a modification of $X_\om$ by an amount $b\mu\ga\De t$. In fact the
transformation (\ref{deltachi}) changes the coefficient $\eta_1$
only, the remaining ones $\eta_2$, $\eta_3$,\dots {\em being left
invariant}. This change exactly cancels the modification of $t$ itself:
\begin{equation}
  \exp\left\{\ga t-{1\over b\mu}X_\om(\ga,\mu)\right\} 
  \to
  \exp\left\{\ga (t + \De t) -{1\over b\mu}(X_\om(\ga,\mu) +
    b\mu\ga \De t)\right\}\,,
\end{equation}
thus implying that the physical results are independent of the
$\La$-parameter choice.  This automatic resummation of the
renormalization scale alleviates the need for techniques such as BLM
resummation~\cite{BLM}, advocated for example in~\cite{BFKLP}, which
show a strong renormalization scheme dependence.

\begin{figure}[htb]
  \begin{center}
    \epsfig{file=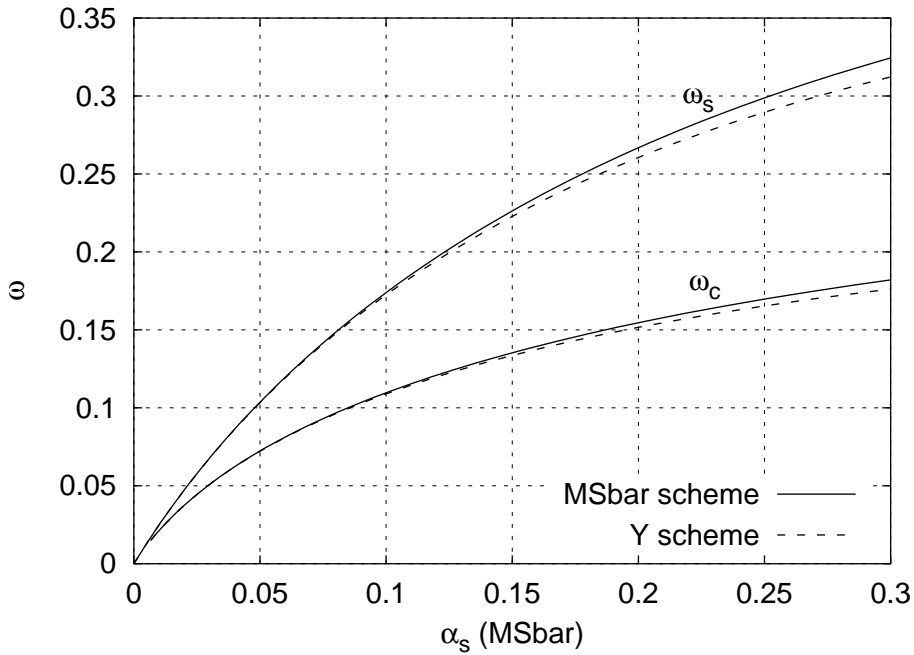}
    \caption{Renormalization scheme uncertainty of the two exponents;
      $\overline{MS}$ scheme and $\Upsilon$ scheme; $\as$ is always
      shown in the $\overline{MS}$ scheme, and is connected to the
      $\Upsilon$ scheme value of $\as$ via \eqref{schmchng}.}
    \label{fig:scheme}
  \end{center}
\end{figure}

The issue of renormalization scheme dependence is in fact closely related.
Consider a scheme $S$ related to the $\overline{MS}$ scheme by 
\begin{equation}
  \as^{(S)} = \as^{(\MSbar)} + T \as^2\,,
  \label{schmchng}
\end{equation}
with an appropriate modification of $\chi_1^\om$.  Except for terms of
$\cO{\as^3}$ and higher, this is identical to a renormalization scale
change. Indeed if one defines the scheme change by a modification of
$\Lambda$ then renormalization scheme changes behave exactly as
renormalization scale changes, and so have no effect on the answer.
Using instead \eqref{schmchng} there is some residual dependence on
the scheme at $\cO{\as^3}$, but as one can see in
Fig.~\ref{fig:scheme} for the $\Upsilon$ scheme, which has
$T\simeq1.17$ (for $\nf=0$), the effect of the change of scheme is
small.
%This is actually true only insofar as the renormalization scheme
%change itself is well defined. Just as valid as \eqref{schmchng} is 
%\begin{equation}
%  \as^{(\MSbar)} = \as^{(S)} - T \as^2\,,
%  \label{schmchngb}
%\end{equation}
%which differs at $\cO{\as^3}$. But with the latter (for positive
%$T$) the largest possible value of $\as^{(\MSbar)}$ is $1/4T$. 
%, and while for positive $T$,
%\eqref{schmchng} is fairly stable, \eqref{schmchngb}  is less
%so.  Inverting   \eqref{schmchngb} gives
%\begin{equation}
%  \as^{(S)} = \frac{1 - \sqrt{1 - 4T\as^{(\MSbar)}}}{2T}\,,
%  \label{schmchngbinv}
%\end{equation}
%and it is evident there are problems inherent to the scheme change
%(and not to the small-$x$ problem) for $\as^{(\MSbar)} > 1/4T$. It
%should be borne in mind that $T$ can quite easily be of order $1$.
%\textbf{DO WE REALLY WANT TO DISCUSS ALL THIS?}

\begin{figure}[htbp]
  \begin{center}
    \epsfig{file=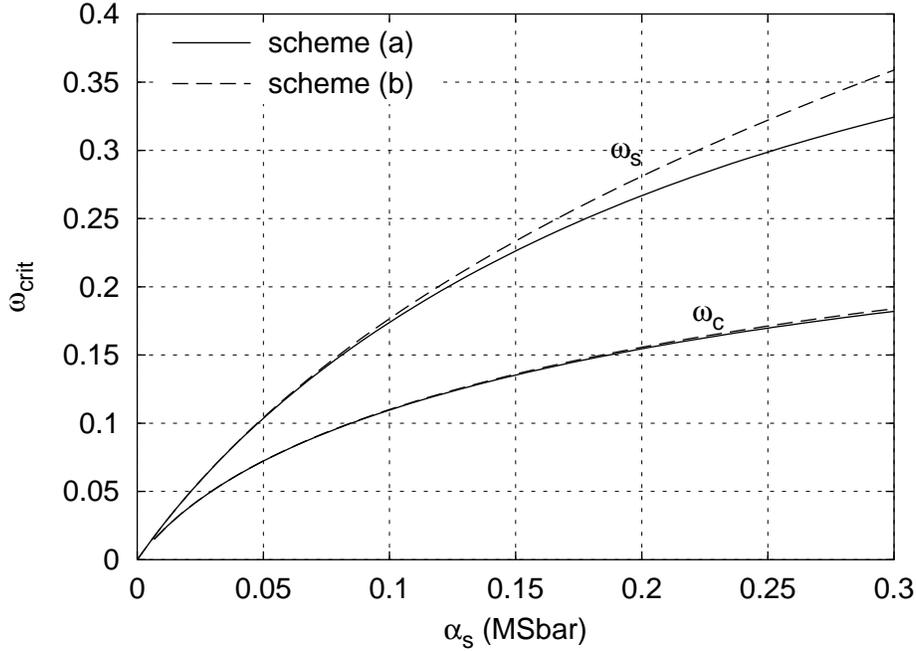}
    \caption{Resummation scheme uncertainty of the two exponents.}
    \label{fig:resumscheme}
  \end{center}
\end{figure}

\paragraph{Resummation scheme.}
In resumming the double transverse logarithms (energy-scale terms),
there is some freedom in one's choice of how to shift the poles around
$\ga=0$ and $\ga=1$. In a similar manner to what was done
in~\cite{s98} we consider two choices. The one explicitly discussed in
this paper (and the one used for all the figures elsewhere in this
paper) can be summarized as
\begin{equation}
  \psi^{(n-1)}(\ga) \to \psi^{(n-1)}(\ga + \ho)\,,
\end{equation}
with an equivalent procedure around $\ga=1$. We refer to this as
resummation type (a).  An alternative possibility is
\begin{equation}
  \frac{1}{\ga^n} \to \frac{1}{(\ga + \ho)^n}\,.
\end{equation}
Thus we have
\begin{subequations}
\begin{align}
  \chi_0^\om(\ga) &= \chi_0(\ga) - \frac1{\ga} - \frac1{1-\ga}
  + \frac1{\ga+\ho} + \frac1{1-\ga+\ho}\,,\\
  \chi_1^\om(\ga) &= \tilde{\chi}_1(\ga)+\frac{A_1(\om)}{(\ga+\ho)^2}+
  \frac{A_1(\om)-b}{(1-\ga+\ho)^2}+\ds\frac12\left(\frac1{\ga+\ho} +
    \frac1{1-\ga+\ho}\right)\,,\\
  \tilde{\chi}_1(\ga)&=\chi_1(\ga)+{\chi_0(\ga)\over2}
  \left(\frac1{\ga^2} +
    \frac1{(1-\ga)^2}\right)-{1\over2}\left(\frac1{\ga} +
    \frac1{1-\ga}\right)-\frac{A_1(0)}{\ga^2}
  -\frac{A_1(0)-b}{(1-\ga)^2}\,.
\end{align}
\end{subequations}
A comparison of these two resummation schemes is given in
Fig.~\ref{fig:resumscheme} and the difference between them is again
reasonably small. 

Aside from the explicit renormalization-scale independence, the
stability of our approach is connected with the resummation of the
collinear poles, for both the double-log, energy-scale dependent terms
(the $1/\ga^3$ and $1/(1-\ga)^3$ poles at NLO) and for the single-log
ones of Eqs.~(\ref{1loop}) and (\ref{mueff}).
%The resummation is such that a
%modification at (say) NNL gets compensated by a change in the
%energy-scale terms at higher orders. Thus the results here for
%$\om_s$, with the proper inclusion of the DGLAP and running-$\as$
%collinear terms, differ little from those in \cite{s98}.
Stability has been noted elsewhere, in the study of a rapidity veto
(initially examined in \cite{schm99}) combined with a resummation of
the energy-scale terms \cite{frs99}.

%======================== 6 ============================

\section{Conclusions}\label{sec:concl}

In this paper, we have improved the small-$x$ equation in several
ways. Firstly, we have taken into account the collinear limits, and
their scale dependence. This implies the $\om$-shifts of the
$\ga$-singularities in Eq.~(\ref{chin}), which yield a double-log
resummation of parameters like $\om/\ga\simeq\as/\ga^2$ or
$\om/1-\ga$, and implies also the effective characteristic function in 
Eq.~(\ref{mueff}), which yields a single-log resummation in the parameter
$\as\chi_1/\chi_0\simeq\as/\ga$ or $\as/1-\ga$.

Both kinds of resummation require an infinite number of subleading
terms in the original BFKL formalism, in which both $\om$ and the
running coupling play the role of expansion parameters. The RG
improved kernel (Eq.~(\ref{seriet})) is actually an infinite series in
$\as(t)$ of $\om$-dependent kernels, so that the corresponding
Eq.~(\ref{eqbfkl}) is no longer an evolution equation in $\log1/x$ with a
simple dependence on the conjugate variable $\om$, but a much more
general $\om$-dependent integral equation.

The second important improvement concerns the treatment of this
generalized equation. In the limit in which the Green's function is
factorized (Eq.~(\ref{fattg})) we have singled out the solutions of
the homogeneous equation $\F_\om(\kk)$ ($\tilde{\F}_\om(\kk_0)$) which
are regular for $t\to+\infty$ ($t\to-\infty$), and we have provided a
general method for the construction of $\F_\om(\kk)$ in
Eq.~(\ref{rappres}). The latter exploits $\om$ as expansion parameter
(or the eigenvalue $\mu\neq\om$ if referred to the eigenfunctions
(\ref{autof})), and thus we call it $\om$-expansion. It allows the
construction, in terms of the improved kernels, of the characteristic
function $\chi(\ga,\om)$ of Eqs.~(\ref{rapfin}) and (\ref{ps}), which
shows no sign of instability when $\om$ increases
(Fig.~\ref{fig:chigaom}), even if the improved kernel is truncated at
NL accuracy.

The key advantages of the improved equation concern the calculation of 
the $t$-de\-pen\-dence, or of the resummed anomalous dimensions, which can 
be given in terms of $\F_\om(\kk)$ only. Let us list some of them:
\begin{itemize}
\item The resummation involves not only all powers of $\as/\om$,
but also an infinite number of subleading terms and extrapolates quite 
smoothly the fixed order perturbative result (Fig.~\ref{fig:anomdim}).
\item Although we resum only a fraction of such subleading terms, we
have characterized the error that we make as a constant scale change
$\De t=\ord(\om)$, or $\De\as/\as=\ord(\as\,\om)$. Therefore, the
neglected terms are subleading, order by order, in both $\as/\om$ and
$\as$ expansions.
\item Although we are limited in principle to small $\om$'s, we
incorporate exact one-loop (and partly two-loop) anomalous dimensions
in the $\om$-dependent kernels. In particular, we have exact
energy-momentum conservation, i.e., the gluon anomalous dimension
vanishes for $\om=1$.
\end{itemize}
We have provided two critical exponents $\om_s(t)$ and
$\om_c(t)$ that signal the breakdown of the above resummation. The
first one ($\om_s(t)$) is roughly related to the breakdown of the
$\as/\om$ resummation, or better of the saddle point
(``semiclassical'') approximation, valid for large $b\om t$ (or
$\as/\om\ll1$), and was the only one considered in previous L+NL
estimates. The latter exponent ($\om_c(t)$) comes from a zero of the
gluon density and signals a singularity of the resummed anomalous
dimension series. Their difference involves non-integer powers of
$\as$ ($\as^{5/3}$ and higher) which are related to a ``quantum''
wavelength in the $t$-dependence.

The estimates of $\om_s(t)$ and $\om_c(t)$ (Fig.~\ref{fig:omcoms}) in
the improved formulation are now quite stable (Figs.~\ref{fig:scheme}
and \ref{fig:resumscheme}) --- despite the large size of NL
corrections --- and nearly re\-nor\-ma\-li\-za\-tion-scheme
independent! The reason for that stems from both the collinear
improvement of the kernel, and from the RG invariant formulation of
the solution. Both exponents are actually useful for a full
understanding of the solution $\tilde{\F}_\om(\kk_0)$, carrying the
(non-perturbative) pomeron singularity $\op$. Indeed we have argued
that --- for reasonable strong coupling extrapolations (positive
definite $\as(t)$) --- the pomeron intercept is bounded between
$\underset{t}{\max}\om_c(t)$ and $\underset{t}{\max}\om_s(t)$. Present
estimates of the latter (Fig.~\ref{fig:omcoms}) are consistent with
the small-$x$ exponent $\simeq0.2$ seen for moderate $Q^2$ at
HERA~\cite{hera}. But a detailed analysis, including two-scale
processes~\cite{2dis,fjet}, is required to obtain a clearcut picture. 

%From the present estimates (Fig.~\ref{fig:omcoms}) we can perhaps risk the
%statement that $0.2\leq\op\leq0.3$, which could be reasonable compared
%to HERA data~\cite{hera}.

Having no problems with stability, we are now more confident of future
progress. We have already mentioned the need for evaluating
$\tilde{\F}_\om(\kk)$, the regular solution for $t\to-\infty$, which
is much more dependent on the strong coupling region. But also the
full Green's function $\G_\om(\kk,\kk_0)$ for $k/k_0=\ord(1)$ ---
i.e., outside the the factorization regime --- is interesting for the
description of two-scale processes (double DIS~\cite{2dis}, forward
jet~\cite{fjet}, etc.). We hope to have a better understanding of both
quantities from a simple model with collinear
resummation~\cite{cds99b}.

Of course, a complete understanding involves a variety of other
questions, like a realistic evaluation of $\op$, a full inclusion of
quarks (Sec.~\ref{ss:qq}) and impact factors~\cite{cd98,bsh97},
the relation to the CCFM equation~\cite{c88,cfm90} and other two-channel
formulations~\cite{kms97}, and so on. But we think that, despite some
residual uncertainties, we are on the right track.

%===================================================

\subsection*{Acknowledgments}
We wish to thank Yuri Dokshitzer, Jeffrey Forshaw, Peter Landshoff,
Pino Marchesini, Douglas Ross and Robert Thorne for helpful
discussions.

%================    APPENDICE   ====================

\appendix

\section{$\mu$-Expansion}

\subsection{Saddle-point method}\label{a:puntosella}

In this appendix we want to show the saddle point procedure for
deriving the dependence of the coefficients $\eta_j^\om:j=0,1,\dots$ of the
small-$\mu$ expansion for $\chi_\om(\ga,\mu)$
\begin{equation}
  \chi_{\om}(\ga,\mu)=\eta^{\om}_0(\ga)+\mu\,\eta_1^{\om}(\ga)+\mu^2
 \,\eta_2^{\om}(\ga)+\cdots\,,     \label{defeta}
\end{equation}
in terms of the eigenvalue functions $\chi^\om_j:j=0,1,\dots$ of the
coefficient kernels in Eq.~(\ref{serie}).

The action of the improved kernel on its eigenfunctions is
\begin{equation}
 0=[\K_\om-\mu]\F_\om^\mu(\kk)={\ab(t)\over\kk^2}\int{\dif\ga\over
 2\pi\ui}\esp{\ga t-{1\over b\mu}X_\om(\ga,\mu)}\left[\sum_{n=0}^{\infty}
 \ab(t)^n\chi_n^\om(\ga)-b\mu t\right]   \label{a1}
\end{equation}   
Now we assume the above integral to be dominated by a saddle point at
$\ga=\gb_\om(\mu,t)$ where $b\mu t=\chi_\om(\ga,\mu)$, $\chi_\om$
being the $\ga$-derivative of $X_\om$ (see Eq.~(\ref{puntos})). By
adopting $\gb$ and $\mu$ as independent variables, we replace
\begin{equation} 
 \ab(t)={1\over bt}={\mu\over\chi_\om(\gb,\mu)}\,.\label{a2}
\end{equation} 
Introducing the ``mean value'' 
\begin{align} 
 \vm{A(\ga)}&\equiv{\int A(\ga)\esp{-V_\om(\ga,\mu)}\;\dif\ga\over
 \int\esp{-V_\om(\ga,\mu)}\;\dif\ga}\,,            \label{vm}\\
 -V_\om(\ga,\mu)&\equiv\left[\ga t-{1\over b\mu}X_\om(\ga,\mu)\right]
 -[\ga\to\gb]=-{1\over b\mu}\sum_{m=2}^\infty{1\over m!}\;
 \chi^{(m-1)}_\om(\gb,\mu)\,\De^m\,,                    \label{poteff}
\end{align}
where ${\ds\chi^{(m)}_\om}\equiv\de_\ga^m\chi_\om$ and
$\De\equiv\ga-\gb$, we write Eq.~(\ref{a1}) in the form
\begin{align}
 \chi_\om(\gb,\mu)=b\mu t=\sum_{n=0}^{\infty}\left[{\mu\over
 \chi(\gb,\mu)}\right]^n\vm{\chi_n(\ga)}\,,        \label{a3}
\end{align}
having dropped the $\om$ dependence. We observe that
\begin{align}
 0&=\left(\lim_{\ga\to+\ui\infty}-\lim_{\ga\to-\ui\infty}\right)\esp{-V}=
 \int\dif\ga\;\de_\ga\left(\esp{-V}\right)
 \propto\vm{\de_\ga V}={1\over b\mu}
 \big[\vm{\chi(\ga,\mu)}-b\mu t\big]\,.           \label{a4}
\end{align}
Collecting Eqs.~(\ref{a1}), (\ref{a3}) and (\ref{a4}) we obtain the
basic equation
\begin{equation}
 \vm{\chi(\ga,\mu)}=\sum_{j=0}^\infty\mu^j\vm{\eta_j(\ga)}=
 \sum_{n=0}^{\infty}\left[{\mu\over\chi(\gb,\mu)}\right]^n
 \vm{\chi_n(\ga)}\,.        \label{eqbase}
\end{equation}
At lowest order ($\ord(\mu^0)$) we have
\begin{equation}
 \vm{\eta_0(\ga)}=\vm{\chi_0(\ga)}\,.      \label{eta0}
\end{equation}
One can easily check that\footnote{We denote as $[x]$ the integer
part of $x$.} $\vm{\De^n}=\ord(\mu^{\left[{n+1\over2}\right]})$.
Since
\begin{equation}
 \vm{A(\ga)}=\vm{A(\gb)}+\vm{A'(\gb)\,\De}+\cdots=A(\gb)+\ord(\mu)
\end{equation}
it follows that, for all $\gb$, $\eta_0(\gb)=\chi_0(\gb)$ and hence
$\eta_0=\chi_0$. Taking into account Eq.~(\ref{eta0}), we can simplify
Eq.~(\ref{eqbase}) as
\begin{equation}
 \sum_{j=1}^\infty\mu^{j-1}\vm{\eta_j(\ga)}=
 \sum_{n=1}^{\infty}{\mu^{n-1}\over[\chi(\gb,\mu)]^n}
 \vm{\chi_n(\ga)}\,.        \label{eqsemp}
\end{equation}
The lowest order of this new relation yields
\begin{equation}
 \vm{\eta_1(\ga)}={\vm{\chi_1(\ga)}\over\chi(\gb,\mu)}={\chi_1(\gb)\over
 \eta_0(\gb)}+\ord(\mu)={\chi_1(\gb)\over\chi_0(\gb)}+\ord(\mu)
\end{equation}
and hence $\eta_1=\chi_1/\chi_0$. The next order reads
\begin{equation}
 \vm{\eta_1(\ga)}+\mu\vm{\eta_2(\ga)}={\vm{\chi_1(\ga)}\over
 \chi(\gb,\mu)}+\mu\,{\vm{\chi_2(\ga)}\over[\chi(\gb,\mu)]^2}\,.
\end{equation}
By expanding with respect to $\ga$ around $\gb$ yields
\begin{align}
 \eta_1(\gb)&+\eta'_1(\gb)\vm{\De}+{1\over2}\eta''_1(\gb)\vm{\De^2}+
 \mu\,\eta_2(\gb)=\nonumber\\
 &={1\over\chi(\gb,\mu)}\left[\chi_1(\gb)+\chi'_1(\gb)\vm{\De}+
 {1\over2}\chi''_1(\gb)\vm{\De^2}\right]+\mu\,{\chi_2(\gb)\over[
 \chi(\gb,\mu)]^2}\,.      \label{ord2}
\end{align}
To the relevant order in $\mu$, we have
\begin{equation}
 \vm{\De}=-{b\mu\over2}\,{\chi''_0(\gb)\over[\chi'_0(\gb)]^2}\,,
 \qquad\vm{\De^2}={b\mu\over\chi'_0(\gb)}\,,\qquad
 {1\over\chi(\ga,\mu)}={1\over\chi_0(\gb)}\left(1-\mu\,{\eta_1(\gb)\over
 \chi_0(\gb)}\right)\,,
\end{equation}
and substituting in Eq.~(\ref{ord2}) we get
\begin{align*}
 \left(\eta_2-{\chi_2\over\chi_0^2}+{\eta_1^2\over\chi_0}\right)
 &=-{\chi'_0\chi_1
 \over\chi_0^2}{b\mu\over2}{\chi''_0\over[\chi'_0]^2}+
 {b\mu\over2\chi'_0}\left({\chi''_0\chi_1\over\chi_0^2}+2{\chi'_0
 \chi_1\over\chi_0^2}-2{[\chi'_0]^2\chi_1\over\chi_0^3}\right)\\
 &={b\over\chi_0}\,\eta'_1\,,
\end{align*}
i.e.,
\begin{equation}
 \eta_2={1\over\chi_0}\left[{\chi_2\over\chi_0}-(\eta_1-b\de_\ga)\eta_1
 \right]\,.
\end{equation}
Going further requires taking into account higher order terms both in
the fluctuations $\vm{\De^m}$ and in the $\mu$-expansion of
Eq.~(\ref{eqsemp}).

The advantage of this method is that it is clearly local in $t$
(because the saddle point $\gb_\om$ is a function of $t$) and in $\ga$ 
(because of the finite fluctuations). Therefore, if $t$ is large
enough for a stable saddle point to exist, then the procedure and the
result are independent of the regularization procedure in the strong
coupling region $t\simeq0$.

The disadvantage, though, is that the order of fluctuations required
increases rapidly with the $\mu$-exponent. It turns out in fact that,
in order to determine $\eta_j:j>2$ --- i.e., to evaluate
Eq.~(\ref{eqsemp}) to order $\mu^{j-1}$ ---, the most involved
calculation concerns $\vm{\De}$ which requires the computation of the
fluctuations in $\int\De\,\esp{-V}\,\dif\ga$ up to order $6j-8$.

\subsection{$\ga$-Derivative method}\label{a:dergamma}

By comparison, the $\ga$-derivative method is formally much
simpler. We start rewriting the basic equation~(\ref{equazdiff}) by
introducing the notation
\begin{align}
 &\chi_\om(\ga,\mu)-\chi_0(\ga)\equiv\mu\,\eta_\om(\ga,\mu)\,,\qquad
 \eta_\om(\ga,\mu)=\eta_1^\om(\ga)+\mu\,\eta_2^\om(\ga)+\mu^2\,
 \eta_3^\om(\ga)+\cdots\,,\nonumber\\
 &D\equiv(\eta_\om-b\,\de_\ga)
\end{align}
in the form
\begin{align}
 \eta_\om(\ga,\mu)&=\left[\chi_0^\om+\mu D\right]^{-1}\chi_1^\om+
 \left[\chi_0^\om+\mu D\right]^{-2}\mu\,\chi_2^\om+\cdots\nonumber\\
 &=\left(1+{\mu\over\chi_\om}\,D\right)^{-1}\left({\chi_1^\om\over
 \chi_0^\om}+\chi_0^\om{}^{-1}\left(1+{\mu\over\chi_\om}\,D
 \right)^{-1}\mu{\chi_2^\om\over\chi_0^\om}+\cdots\right)\,.\label{oprt}
\end{align}
We then expand in $\mu$ both the $\chi_n^\om$ series and the operator
denominators, which depend on $\eta_\om$ in a non-linear way, and we
derive the $\mu$-expansion~(\ref{omegaes}).

For instance, if we want $\eta_3^\om$, we can rewrite Eq.~(\ref{oprt}) 
up to order $\mu^2$ in the form
\begin{align}
 \eta&=\left(1-\mu\,\chi_0^{-1}D+\mu^2(\chi_0^{-1}D)^2\right){\chi_1
 \over\chi_0}+{1\over\chi_0}\left(1-\mu(D\chi_0^{-1}+\chi_0^{-1}D)
 \right)\mu{\chi_2\over\chi_0}+\mu^2{\chi_3\over\chi_0}\nonumber\\
 D&=D_1+\mu\,\eta_2+\cdots\,,\qquad D_1\equiv\eta_1-b\,\de_\ga\,,
 \label{etacoef}
\end{align}
where the $\om$ index has been dropped.

We then identify the $\eta_i$ coefficients in Eq.~(\ref{etacoef}) term 
by term:
\begin{align}
 \eta_1&={\ds\chi_1\over\chi_0}\,,\qquad\eta_2={1\over\chi_0}\left(
 {\chi_2\over\chi_0}-D_1{\chi_1\over\chi_0}\right)\,,\label{riete}\\
 \eta_3&={\chi_3\over\chi_0}-{1\over\chi_0}\left(D_1{1\over\chi_0}+
 {1\over\chi_0}D_1\right){\chi_2\over\chi_0}+\left({1\over\chi_0}D_1
 \right)^2{\chi_1\over\chi_0}-{1\over\chi_0}\eta_2{\chi_1\over\chi_0}
 \,,\nonumber
\end{align}
Eq.~(\ref{riete}) proves Eq.~(\ref{ete}) of the text.

We notice the curious fact that if
\begin{equation}
 {\chi_2\over\chi_0}=D_1{\chi_1\over\chi_0}\,,\qquad
 {\chi_3\over\chi_0}=D_1^2{\chi_1\over\chi_0}\,,\label{curioso}
\end{equation}
both $\eta_2$ and $\eta_3$ vanish identically. This is a particular
case of Eq.~(\ref{ricorrenza}), which states that
$\eta_1=\chi_1/\chi_0$ is an exact solution of Eq.~(\ref{oprt}) if
\begin{equation}
 {\chi_{n+1}\over\chi_0}=D_1^n{\chi_1\over\chi_0}\,,\qquad n\geq1
 \,.   \label{chinpu}
\end{equation}
In fact we have the chain of identities
\begin{align}
 {\chi_1\over\chi_0}&=\sum_{n=0}^\infty\left[(\chi_0+\mu D_1)^{-n}
 (\mu D_1)^n-(\chi_0+\mu D_1)^{-(n+1)}(\mu D_1)^{n+1}\right]
 {\chi_1\over\chi_0}\label{catena}\\
 &=\sum_{n=0}^\infty(\chi_0+\mu D_1)^{-(n+1)}\chi_0(\mu D_1)^n
 {\chi_1\over\chi_0}=\sum_{n=0}^\infty(\chi_0+\mu D_1)^{-(n+1)}
 \chi_{n+1}\,,
\end{align}
which prove Eq.~(\ref{oprt}) if Eq.~(\ref{chinpu}) is satisfied.

It is straightforward to check that the ansatz (\ref{chinpu}) builds
up the correct collinear singularities to all orders. We start from
\begin{equation}
 {\chi_1\over\chi_0}\simeq{A_1\over\ga}\quad,\quad{A_1-b\over1-\ga}
 \qquad(\ga\to0,1)\label{parto}
\end{equation}
and, by applying the $D_1$ operator of Eq.~(\ref{chinpu}) in the
relevant limits we obtain the result
\begin{equation}
 {\chi_{n+1}\over\chi_0}\simeq\left({A_1\over\ga}-b\,\de_\ga\right)^n
 {A_1\over\ga}\quad,\quad\left({A_1-b\over1-\ga}-b\,\de_\ga\right)^n
 {A_1-b\over1-\ga}\,,\label{fine}
\end{equation}
which checks with Eq.~(\ref{chin}). It follows that the leading collinear
singularities must cancel out in $\eta_j:j\geq2$, as stated in
Sec.~\ref{ss:eoe}.

We should keep in mind that the two methods just illustrated are
equivalent when both make sense, i.e., for $t$ large enough for the
stable saddle point to exist. This assumption is implicitly present in
the $\ga$-derivative method when we expand the operators in
Eq.~(\ref{oprt}). This means that we stay away from the zero modes of
the full operator and we consider the $D$ operator as a small
perturbation with respect to $\chi_0$. Expanding in $D$ is analogous
to the fluctuation expansion.

%===================================================


\begin{thebibliography}{99}
\bibitem{fl98} V.S. Fadin and L.N. Lipatov,\plb{429}{127}{1998}
                     and references therein.
\bibitem{cc98} G. Camici and M. Ciafaloni,\plb{412}{396}{1997};\\
                      G. Camici and M. Ciafaloni,\plb{430}{349}{1998}.
\bibitem{bfkl}
    E.A. Kuraev, L.N. Lipatov and V.S. Fadin,\spj{45}{199}{1978};\\
    Y.Y. Balitski and L.N. Lipatov,\sjnp{28}{22}{1978}.
\bibitem{dglap}
    V.N. Gribov and L.N. Lipatov,\sjnp{15}{438}{1972};\\
    G. Altarelli and G. Parisi,\npb{126}{298}{1977};\\
    Y.L. Dokshitzer,\spj{46}{641}{1977}.
\bibitem{c88} M. Ciafaloni,\npb{296}{49}{1988}.
\bibitem{cfm90}
    S. Catani, F. Fiorani and G. Marchesini,\plb{234}{339}{1990};\\
    S. Catani, F. Fiorani and G. Marchesini,\npb{336}{18}{1990}.
\bibitem{kms97}
         J. Kwiecinski, A.D. Martin and A.M. Stasto,\pr{56}{3991}{1997};
see also\\M. Krawczyk, {\it Nucl. Phys. Proc. Suppl.} {\bf 18C} (1991) 64.
\bibitem{cd99} M. Ciafaloni and D. Colferai,\plb{452}{372}{1999}.
\bibitem{s98} G.P. Salam, \jhep{9807}{19}{1998}.
\bibitem{ck89}J. Kwiecinski,\zp{29}{561}{1985};\\
                      J.C. Collins and J. Kwiecinski,\npb{316}{307}{1989}.
\bibitem{cc97} G. Camici and M. Ciafaloni,\plb{395}{118}{1997}.
\bibitem{glr}
    L.V. Gribov, E.M. Levin and M.G. Ryskin,\prep{100}{1}{1983}.
\bibitem{l86} L.N. Lipatov,\spj{63}{904}{1986}.
\bibitem{hr92} R.E. Hancock and D.A. Ross,\npb{383}{575}{1992};\\
                      R.E. Hancock and D.A. Ross, \npb{394}{200}{1993}.
\bibitem{nz94} N.N. Nikolaev and B.G. Zakharov,\plb{327}{157}{1994}.
\bibitem{l95} E.M. Levin,\npb{453}{303}{1995}.
\bibitem{b95} M.A. Braun,\plb{345}{155}{1995}.
\bibitem{hkk97}
     L.P.A. Haakman, O.V. Kancheli and J.H. Koch,\plb{391}{157}{1997};\\
     L.P.A. Haakman, O.V. Kancheli and J.H. Koch,\npb{518}{275}{1998}.
\bibitem{km98} Y.V. Kovchegov and A.H. Mueller,\plb{439}{428}{1998}.
\bibitem{bkp80} J. Bartels,\npb{175}{365}{1980};\\
             J. Kwiecinski and M. Praszalowicz,\plb{94}{413}{1980};\\
             L.N. Lipatov,\spj{59}{571}{1994};\\
             A.H. Mueller,\npb{437}{107}{1995}.
\bibitem{c98} M. Ciafaloni,\plb{429}{363}{1998}.
\bibitem{cd98} M. Ciafaloni and D. Colferai,\npb{538}{187}{1999}.
\bibitem{cds99b} M. Ciafaloni, D. Colferai and G.P. Salam, to appear.
\bibitem{dmw96}
      Y.L. Dokshitzer, G. Marchesini and B.R. Webber,\npb{469}{93}{1996}.
\bibitem{cc97b} G. Camici and M. Ciafaloni,\npb{496}{305}{1997}.
\bibitem{bv98} J. Bl\"umlein and A. Vogt,\pr{58}{014020}{1998}.
\bibitem{r98}D.A. Ross,\plb{431}{161}{1998};\\
            E.M. Levin,\hep{9806228}.
\bibitem{t99} R.S. Thorne,\hep{9901331}.
\bibitem{ags96}
       B. Andersson, G. Gustafson and J. Samuelsson,\npb{467}{443}{1996}.
\bibitem{c98b} M. Ciafaloni, Communication at the Durham workshop
                       (1998). 
\bibitem{t65} M. Toller,\ncim{37 n.2}{631}{1965}.
\bibitem{cc96} G. Camici and M. Ciafaloni,\plb{386}{341}{1996}.
\bibitem{BLM} S.J. Brodsky, G.P. Lepage and P.N. Mackenzie,
                     \pr{28}{228}{1983}.
\bibitem{BFKLP} S.J. Brodsky, V.S. Fadin, V.T. Kim, L.N. Lipatov and
                       G.B. Pivovarov,\hep{9901229}.
\bibitem{FRBook} J.R. Forshaw and D.A. Ross, {\em Perturbative QCD
    and the Pomeron}, Cambridge University Press, 1997.
\bibitem{ehw95}
       R.K. Ellis, F. Hautmann and B. R. Webber,\plb{348}{582}{1995}.
\bibitem{bf96} R.D. Ball and S. Forte, {\it Proceedings of the DIS 96
             Workshop}, 1996.
\bibitem{schm99} C. Schmidt, preprint MSUHEP-90122,\hep{9901397}.
\bibitem{frs99}  J.R. Forshaw, D.A. Ross and A. Sabio Vera, preprint
            CERN-TH/99-64, MC-TH-99-04, SHEP -99/02,\hep{9903390}. 
\bibitem{hera}  H1 Collaboration (Adloff et al.),\zp{72}{593}{1996};\\
             ZEUS Collaboration (Breitweg et al.),\plb{407}{402}{1997}.
\bibitem{2dis}  L3 Collaboration (M. Acciarri et al.),\plb{453}{333}{1999}.
\bibitem{fjet} ZEUS Collaboration (J. Breitweg et al.),\epj{6}{239}{1999};\\ 
                    H1 Collaboration (C. Adloff et al.),\npb{538}{3}{1999}.
\bibitem{bsh97}
            S.J. Brodsky, F. Hautmann and D.E. Soper,\pr{56}{6957}{1997}.
\end{thebibliography}
\end{document}